\documentclass[reprint]{revtex4-2}
\usepackage[margin=1in]{geometry} 
\usepackage{amsfonts,amsmath,amsthm,amssymb}
\usepackage{mathtools}
\usepackage{enumitem}
\usepackage{color}
\usepackage{mathrsfs}
\usepackage{setspace}
\usepackage{graphicx}
\usepackage{caption}
\usepackage{subcaption}
\usepackage{hyperref}
\usepackage{bm}
\usepackage{physics}
\usepackage{dcolumn}
\usepackage{todonotes}
\usepackage{nomencl}
\makenomenclature

\graphicspath{./images/}

\DeclareMathOperator*{\argmin}{argmin}

\begin{document}
\pagestyle{plain}
\title{On the non-uniqueness of the quasi-static magnetic forward problem and its impact on source current reconstruction}

\author{Wan-Jin Yeo$^{1,2}$, Yao-Rui Yeo$^{3}$, Aaron Miller$^{1}$, Samu Taulu$^{1,2}$}
\affiliation{$^1$Department of Physics, University of Washington, Seattle, WA 98195, USA \\
$^2$Institute for Learning and Brain Sciences, University of Washington, Seattle, WA 98195, USA \\
$^3$Department of Population Health, New York University, New York, NY 10016, USA}

\date{\today}

\begin{abstract}

We introduce a formulation where individual line segments of a current loop have translationally non-invariant contributions to the electro-quasi-static magnetic scalar potential and magnetic field in source-free regions. While closed current loops composed of these open segments have translationally invariant contributions, our formulation indicates that there are multiple ways to decompose the magnetic field due to a closed current loop into the open current  segments. By defining the path-independent magnetic scalar potential using a radial integration path with respect to a given origin, a formula has been derived that shows that only non-radial line current segments have non-zero contributions, whereas radial current segments are magnetically silent. This indicates that magnetic forward models for open current segments are non-unique, since the orientations of the currents are dependent on the choice of the computational origin. This finding affects the conventional physical interpretation of the significance of primary/volume currents in biomagnetic forward and inverse models of brain activity, since their contributions are shown to vary with the choice of origin in our formulation. As an example, we perform a simple magnetoencephalography (MEG) equivalent current dipole (ECD) fit for a spherical head model with various origin choices to localize a primary current source. The primary current can be made to have zero contributions in certain translated origin choices (and the volume currents contribute to the signal entirely). Thus, there exist origins in which the ECD model fits the data with high confidence but incorrect estimated location of the primary current segment. In such cases, the dipole is fit to a segment of the volume current. If one were unaware of any origin translations, such ECD fits may be incorrectly interpreted as the primary current source. 

\end{abstract}

\maketitle

\nomenclature{\(\mathbf{r}\)}{Field point}
\nomenclature{\(\mathbf{r}'(t)\)}{Source point}
\nomenclature{\(\bm{\gamma}\)}{Lead field (dipolar source)}
\nomenclature{\(\bm{\alpha}\)}{Dipole strength vector}
\nomenclature{\(\mathbf{b}(\mathbf{r})\)}{Magnetic field (open current segment)}
\nomenclature{\(u(\mathbf{r})\)}{Magnetic scalar potential (open current segment)}
\nomenclature{\(\mathbf{B}(\mathbf{r})\)}{Magnetic field (closed current loop)}
\nomenclature{\(U(\mathbf{r})\)}{Magnetic scalar potential (closed current loop)}
\nomenclature{\(\mathbf{a}\)}{Source-to-field vector, $\mathbf{r} - \mathbf{r}'$}
\nomenclature{\(\tilde{\Box}\)}{Quantity associated with a translation by $\mathbf{r}_t$}
\nomenclature{\(ECD\)}{Equivalent current dipole}
\nomenclature{\(k\)}{Radial projection of $\mathbf{a}$, $\mathbf{a}\cdot\mathbf{e}_r$}
\nomenclature{\(I\)}{Magnitude of electric current}
\nomenclature{\(GOF\)}{Goodness of fit}
\nomenclature{\(LE\)}{Localization error}
\nomenclature{\(\mathbf{J}\)}{Total current density}
\nomenclature{\(\Box_{pri}\)}{Quantity associated with primary current}
\nomenclature{\(\Box_{vol}\)}{Quantity associated with volume current}
\nomenclature{\(\Box_{D}\)}{Quantity associated with dipolar source}
\nomenclature{\(\mathbf{D}\)}{Dipole moment}
\nomenclature{\(\mathbf{E}(\mathbf{r})\)}{Electric field}
\nomenclature{\(V(\mathbf{r})\)}{Electric scalar potential}
\nomenclature{\(d\mathbf{L}'\)}{Source space differential, $ -I(\mathbf{r}' \times d\mathbf{l}')$}
\nomenclature{\(d\mathbf{l}'\)}{Differential along current source}
\nomenclature{\(F\)}{$a(ra + \mathbf{a}\cdot\mathbf{r})$}
\printnomenclature

\section{Introduction}

The magnetic field due to a static configuration of electric current is commonly modeled by considering individual contributions of infinitesimally short segments of current. The total magnetic field is then calculated by taking the linear superposition of these contributions. This is the essence of the Biot-Savart law, which provides a simple mathematical framework for calculating the magnetic field for closed loops of current (see, e.g., \cite{griffiths,jackson1999classical,zangwill2013modern,panofsky2005classical}). While the most physically meaningful elementary source of a magnetic field is the magnetic dipole, i.e., an infinitesimally small loop of current, assigning characteristic properties to individual segments of closed loops in certain applications may be warranted, such as interpreting the sources of recorded biomagnteic field topographies.

For instance, in magnetoencephalography (MEG), the source models for magnetic forward calculations are typically split into primary currents, which are generated by brain activity, and the volume currents, which passively complete the current loop together with the primary current. The sum of their magnetic field contributions provide us with the total magnetic field and subsequently the MEG sensor signal, which is the magnetic flux through some sensing area or volume depending on the configuration of the MEG sensor. The primary current, when focal, is commonly modeled as an electric current dipole \cite{hamalainen1993magnetoencephalography,ilmoniemi2019brain,ahlfors2019overview} or a straight line current segment that forms a closed current triangle loop with volume current segments \cite{ilmoniemi1985forward,ilmoniemi2009triangle}. Forward models such as the current dipole model are used in solving the inverse problem where the configuration of the electric current that produced the recorded magnetic fields is estimated. The forward model is first parametrized, then its parameter values are optimized by minimizing the error between the model and recorded data via some objective function \cite{hamalainen1993magnetoencephalography}. 

Magnetic inverse models are non-unique because there always exists more than one current distribution that produces the same magnetic field topography outside the region containing the currents \cite{helmholtz1853ueber}. For instance, an arbitrary number of toroidal current loops may be added to any current configuration without changing the magnetic field pattern. Furthermore, a configuration consisting of a radial primary current and the associated volume current does not produce any magnetic field outside of a spherically symmetric conductor \cite{grynszpan1973model,ilmoniemi1985forward,sarvas1987basic}. It should be noted that in this context, the radial direction is fixed with respect to the center of the spherical conductor. Due to the non-uniqueness, a priori assumptions or constraints are  generally required to address the non-uniqueness issue in inverse modeling \cite{dassios2005non,dassios2013definite,fokas1996inversion,fokas2004unique,fokas2009electro,fokas2012electro,van1988beamforming, van1997localization,hamalainen1994interpreting,matsuura1995selective,schmidt1986multiple,mosher1999source,mosher1998recursive,pascual2002standardized}. Such constraints are helpful in numerical stabilization of ill-posed inverse problems as well. 

Sarvas introduced an elegant strategy of calculating the magnetic field due to a single equivalent current dipole outside of a spherically symmetric volume conductor \cite{sarvas1987basic}. The origin of computation was fixed to be the center of the conducting sphere and it was shown that the contribution of all radial current segments vanishes in this computation. This is a celebrated result as it has been shown that in the spherical conductor, all volume currents can be represented by equivalent radial current segments \cite{geselowitz1967bioelectric,ilmoniemi1985forward,ilmoniemi2009triangle}. Thus, it is possible to completely reconstruct the magnetic field by resorting to the parameters of the (non-radial) primary current only. 

In this paper, we show that Sarvas' strategy can be generalized to any current configuration and conductor profile when calculating the electro-quasi-static field in a current-free region. Moreover, Sarvas' results stem from defining the path-independent magnetic scalar potential of the current loop with a radial integration path. We highlight that in a source-free region under the electro-quasi-static approximation, the contributions of individual current segments to the magnetic field depend on the integration path chosen to define the magnetic scalar potential relative to a chosen origin. With a fixed integration path relative to a chosen computational origin, the magnetic contributions are non-invariant with respect to coordinate translations.

We choose to proceed with a radial integration path consistent with Sarvas, since it is the most practical and mathematically simplest choice. In this case, only non-radial current segments have non-zero contributions regardless of the conductor geometry. The total magnetic field of any closed loop of current is, however, invariant with respect to coordinate translations. It is therefore sufficient to only consider non-radial current segments to fully calculate the total magnetic field. For a spherically symmetric conductor then, different choices of origin thus suggest different sensitivities of the MEG signal to current segments. The observation that the magnetic forward problem for open current segments is non-unique can have importance in the implementation of inverse models applied to magnetic field topographies as well as in the interpretation of reconstructed current distributions. For example, such interpretations are commonly made when reconstructing neural currents based on multi-channel MEG measurements. The freedom to choose an origin allows any straight segment of a closed current loop to be forced to be radial or non-radial and hence magnetically non-contributing or contributing respectively. This results in potentially inaccurate inverse models. It is also therefore not necessarily accurate to form static physical interpretations about sources obtained via an inverse method. 


In Section~\ref{sec:theory}, we derive compact equations for the electro-static magnetic scalar potential and magnetic field in a source-free region. These equations indicate nonzero contributions only from non-radial current segments. In Section~\ref{sec:forward}, we extend on this concept by explicitly showing that the contributions from closed current loops are translationally invariant while the contributions from open current segments are non-invariant. In Section~\ref{sec:meg}, we present the usual theoretical frameworks used in MEG, and assert that they may be formulated equivalently to the derivations presented in the preceding sections. In Sections~\ref{sec:ECD}, \ref{sec: verifications}, and \ref{sec:discussion}, we investigate via simulations and discuss the effects of the results found in Sections~\ref{sec:theory} and \ref{sec:forward} for MEG equivalent current dipole (ECD) fits. Finally, in Section~\ref{sec:conclusion}, we conclude our theory and findings.

\section{Theory} \label{sec:theory}

The magnetic field due to a steady current flow with magnitude $I$ along a closed loop $\mathcal{C}$ is given by the Biot-Savart law
\begin{align}
    \mathbf{B} ( \mathbf{r}) = \frac{\mu_0}{4 \pi} \int_\mathcal{C} \frac{I d \mathbf{l}' \times \left(\mathbf{r} - \mathbf{r}'\right)}{\abs{\mathbf{r} - \mathbf{r}'}^3}, \label{biot}
\end{align}
where $\mathbf{r}=(r_x,r_y,r_z)$ is the field point, $\mathbf{r}'=(r'_x,r'_y,r'_z)$ are source points along $\mathcal{C}$, and $\mu_0$ is the vacuum permeability. Under the electro-quasi-static approximation, i.e., setting 
\begin{align}
    \frac{\partial \mathbf{E}}{\partial t} = \mathbf{0},    
\end{align}
the curl of the magnetic field satisfies   
\begin{align} \label{ampere}
    \grad \times \mathbf{B} (\mathbf{r}) = \mathbf{0}
\end{align}
in any source-free region where $\mathbf{J} = 0$ due to Ampere's law. This allows us to write $\mathbf{B}$ as the gradient of a scalar potential $U$,
\begin{align} \label{BgradU}
    \mathbf{B} (\mathbf{r}) = -\mu_0 \grad U (\mathbf{r}).
\end{align}

Note that the Biot-Savart law does not strictly only apply to closed current loops. However, since we are using it simultaneously with Ampere's law, which requires a closed current loop, we have defined the Biot-Savart law \eqref{biot} with a closed loop $\mathcal{C}$ for convenience. We also note that in the Biot-Savart formulation, the total magnetic field is calculated via the principle of superposition of unique contributions from individual current segments forming the current loop.

The additional condition of a curl-less magnetic field \eqref{ampere} allows us to rewrite the Biot-Savart law via the gradient theorem as follows (this is the strategy used by Sarvas in \cite{sarvas1987basic} as well). Assuming $\mathbf{B} (\mathbf{r})$ is smooth (i.e. no change in permeability), \eqref{ampere} indicates that $\mathbf{B} (\mathbf{r})$ is conservative. Moreover, $\mathbf{B} (\mathbf{r})$ is proportional to the gradient of $U (\mathbf{r})$, thus we may arbitrarily choose a path to integrate over it in the gradient theorem to obtain $U(\mathbf{r})$. For convenience, a radial path along $\mathbf{e}_r = \mathbf{r}/||\mathbf{r}||\equiv \mathbf{r}/r$ was chosen. This leads to
\begin{align} 
    U (\mathbf{r}) &= - \int_0^\infty \grad U (\mathbf{r} + t \mathbf{e}_r) \cdot \mathbf{e}_r \ dt \\
    &= \frac{1}{\mu_0} \int_0^\infty \mathbf{B} (\mathbf{r} + t \mathbf{e}_r) \cdot \mathbf{e}_r \ dt. \label{U_integral}
\end{align}
Substituting in Biot-Savart law \eqref{biot} gives 
\begin{align} 
    U (\mathbf{r}) &= \frac{I}{4 \pi} \int_0^\infty \int_\mathcal{C} \frac{d \mathbf{l}' \times \left(\mathbf{r} + t \mathbf{e}_r - \mathbf{r}' \right) \cdot \mathbf{e}_r}{\abs{\mathbf{r} + t \mathbf{e}_r - \mathbf{r}'}^3}  \ dt \nonumber \\ 
    &= \frac{I}{4\pi} \int_0^\infty \int_\mathcal{C} \frac{(\mathbf{r}-\mathbf{r}')\cross \mathbf{e}_r \cdot d\mathbf{l}'}{\abs{\mathbf{r}+t\mathbf{e}_r - \mathbf{r}'}^3} \ dt \label{U_integral1}
\end{align}
where we have applied the scalar triple product. This expression will be useful in showing the invariance of $U(\mathbf{r})$ under coordinate translations later on in Section~\ref{sec:closed_loop}.

Later on in the paper, beginning in Section~\ref{sec:meg}, we will also use the magneto-quasi-static assumption,
\begin{align}
    \frac{\partial \mathbf{B}}{\partial t} = \mathbf{0}.
\end{align}

\subsection{Choice of integration path in defining \texorpdfstring{$U(\mathbf{r})$}{Ur}} \label{path_choice}

We note here the important point that it is possible to define \eqref{U_integral} with any arbitrary path (defined relative to the origin) taken from $\mathbf{r}$ to infinity, instead of the convenient radial direction.

If one chooses a straight line integration path from $\mathbf{r}$ to infinity (i.e, rotating $\mathbf{e}_r$ by a fixed amount) and proceeds with the subsequent derivations, it will lead to the result that current segments that are either parallel to the integration path or the source-to-field direction are magnetically silent. The latter is in agreement with the Biot-Savart law \eqref{biot}, whereas the additional former observation is due to the reduction of degrees of freedom when the electro-quasi-static assumption \eqref{ampere} is held simultaneously with the Biot-Savart law. Selecting a radial integration path is a special case that allows for a purely source space differential $d\mathbf{L}'$ to be defined in the next section, which then leads to our conclusion that all radially-oriented currents segments are magnetically silent with this radial integration path choice.

For non-straight integration paths, the magnetic contributions of current segments based on their orientations become less obvious. In practice, defining the magnetic scalar potential $U(\mathbf{r})$ with a radial integration path is most likely the preferred option due to the ease of interpretation and computation. Thus, we have chosen to proceed with defining the magnetic scalar potential with a radial integration path.

\subsection{The electro-quasi-static magnetic field in a source-free region}

Now, we derive a formula for the magnetic field under electro-quasi-static approximation produced by a closed constant current loop in a source-free region. First, notice that \eqref{U_integral1} may be further written as
\begin{align}
    U(\mathbf{r}) = \frac{I}{4\pi}\int_\mathcal{C}(\mathbf{r}'\times d\mathbf{l}'\cdot \mathbf{e}_{r})\int_{0}^{\infty}\frac{dt}{\abs{\mathbf{r}+t\mathbf{e}_{r}-\mathbf{r}'}^3}. \label{U_pre}
\end{align}
Let $\mathbf{a} = \mathbf{r} - \mathbf{r}'$, $k = \mathbf{a} \cdot \mathbf{e}_r$ and $y = t + k$. Then $dy = dt$ and we have 
\begin{align}
    \int_0^\infty \frac{dt}{\abs{\mathbf{r}+t\mathbf{e}_r - \mathbf{r}'}^3} &= \int_{k}^\infty \frac{dy}{\left( y^2 + a^2 - k^2 \right)^{3/2}} \nonumber \\
    &= \frac{1}{a(a+k)}. \label{integral_result}
\end{align}
Let us now define the quantities
\begin{align}
    F(\mathbf{r},\mathbf{r}') &\equiv a(ra + \mathbf{a} \cdot \mathbf{r}) \label{F} \\
    \grad F(\mathbf{r},\mathbf{r}') &= \left(\frac{a^2}{r} + \frac{\mathbf{a}\cdot\mathbf{r}}{a} + 2a + 2r\right) \mathbf{r} \nonumber \\
    &\quad - \left(a+2r+\frac{\mathbf{a}\cdot\mathbf{r}}{a}\right)\mathbf{r}' \label{grad_F}
\end{align}
which are chosen for consistency of notation with Sarvas \cite{sarvas1987basic}. Notice that \eqref{integral_result} may be written as 
\begin{align}
    \frac{1}{a(a+k)} = \frac{r}{F(\mathbf{r},\mathbf{r}')}.
\end{align}
If we define $d\mathbf{L}' \equiv -I(\mathbf{r}' \times d\mathbf{l}')$, then the scalar potential may  be written as 
\begin{align}
    U(\mathbf{r}) = -\frac{1}{4\pi}\int_\mathcal{C} \frac{d\mathbf{L}'\cdot \mathbf{r}}{F(\mathbf{r},\mathbf{r}')}, \label{eq:novel_U}
\end{align}
and $\mathbf{B}$ can be obtained via \eqref{BgradU},
\begin{align}
    \mathbf{B}(\mathbf{r}) = \frac{\mu_0}{4\pi}\int_\mathcal{C} d\mathbf{L}'\cdot \grad \left[\frac{\mathbf{r}}{F(\mathbf{r},\mathbf{r}')}\right],
\end{align}
where the gradient is taken with respect to $\mathbf{r}$. The gradient portion is a $3 \times 3$ Jacobian matrix, with entries defined as
\begin{align}
    \left\{ \grad \left[\frac{\mathbf{r}}{F(\mathbf{r},\mathbf{r}')} \right]\right\}_{i,j} &= \frac{\partial_{r_j}r_i}{F(\mathbf{r},\mathbf{r}')} + r_i \partial_{r_j} \left(\frac{1}{F(\mathbf{r},\mathbf{r}')}\right) \nonumber \\
    &= \frac{\delta_{ij}}{F(\mathbf{r},\mathbf{r}')} - \frac{r_i \partial_{r_j} F(\mathbf{r},\mathbf{r}')}{F^2(\mathbf{r},\mathbf{r}')} \label{eq:grad_r/F}
\end{align}
where $i,j,=x,y,z$, and $\delta_{ij}$ is the Kronecker delta function. Thus,
\begin{align}
    B_j (\mathbf{r}) &= \frac{\mu_0}{4\pi}\int_\mathcal{C} \sum_{i} dL'_i \left[ \frac{\delta_{ij}}{F(\mathbf{r},\mathbf{r}')} - \frac{r_i \partial_{r_j} F(\mathbf{r},\mathbf{r}')}{F^2(\mathbf{r},\mathbf{r}')}\right] \nonumber
\end{align}
i.e.
\begin{align}
    \mathbf{B} (\mathbf{r}) 
    &= \frac{\mu_0}{4\pi}\int_\mathcal{C} d\mathbf{L}' \cdot \left[  \frac{ \mathbb{I}}{F(\mathbf{r},\mathbf{r}')} - \frac{\mathbf{r} \grad F(\mathbf{r},\mathbf{r}')}{F^2(\mathbf{r},\mathbf{r}')} \right]. \label{eq:novel_B}
\end{align}
As a simple check for correctness, in Appendix~\ref{sec:appendix_dipolecheck}, we show that in the far-field approximation, \eqref{eq:novel_U} and \eqref{eq:novel_B} reduce to the expected expressions for a magnetic dipole.
 
In \eqref{eq:novel_U} and \eqref{eq:novel_B}, the term $d\mathbf{L}'$ is strictly bound to the source space and its value depends on the choice of origin defining the coordinate system. Mathematically, $\mathbf{L}'$ is similar to the angular momentum of the charge carriers divided by their mass, and angular momentum is specific to the choice of origin. Thus, according to \eqref{eq:novel_B}, the contribution of individual infinitesimal segments of the current loop to the total magnetic field will change as we translate the origin because only the non-radial current segments have non-zero contribution to $d\mathbf{L}'$.

If we define a similar quantity $d\mathbf{L}_{f}'=-Id\mathbf{l}'\times(\mathbf{r}-\mathbf{r'})$, which represents angular momentum divided per mass as observed from the field point $\mathbf{r}$, then the Biot-Savart law reads 
\begin{align}
    \mathbf{B} ( \mathbf{r}) = \frac{\mu_0}{4 \pi} \int_\mathcal{C} \frac{d\mathbf{L}_{f}'}{\abs{\mathbf{r} - \mathbf{r}'}^3}.
\end{align}
Our formula, which is consistent with the Biot-Savart formula in a source-less region, differs from Biot-Savart in terms of non-uniqueness of the current segment contributions. This stems from requiring \ref{ampere} to simultaneously hold true with the Biot-Savart formula. In Biot-Savart, the  $d\mathbf{L}_{f}'$ term is fixed to the field point whereas in our formula $d\mathbf{L}'$ is fixed to the computational origin that can be translated without changing the magnetic field at the field point. 

The $d\mathbf{L}'$ in equations \eqref{eq:novel_U} and \eqref{eq:novel_B} indicate that whenever segments of $\mathcal{C}$ have $\mathbf{r}'$ parallel to $d\mathbf{l}'$, then $U(\mathbf{r}) = 0$ and $\mathbf{B}(\mathbf{r}) = \mathbf{0}$. In other words, straight line current segments that are radial with respect to the origin have zero field contributions; only non-radial components of a current loop have nonzero contributions to the $U(\mathbf{r})$ and $\mathbf{B}(\mathbf{r})$ fields. We note that the same conclusion can be reached when using the approach of expanding the magnetic field in vector spherical harmonics \cite{taulu2020unified}.

\section{Effects of origin translations in the magnetic forward problem} \label{sec:forward}

In this section, we first show the expected result that under electro-quasi-static and source-less conditions, the magnetic scalar potential $U(\mathbf{r})$ and magnetic field $\mathbf{B}(\mathbf{r})$ are invariant under coordinate translations when closed current loops are considered. However, individual (open) segments of the source current have non-unique contributions under origin translations.

\subsection{Translational invariance of \texorpdfstring{$U(\mathbf{r})$}{Ur} and \texorpdfstring{$\mathbf{B}(\mathbf{r})$}{Br} for closed current loops} \label{sec:closed_loop}

First, we show that \eqref{U_integral1}, which is the magnetic scalar potential $U(\mathbf{r})$ due to a closed current loop, remains invariant under coordinate translations. If we translate the origin by $\mathbf{r}_t$, i.e.
\begin{align}
    \mathbf{r} &\rightarrow \tilde{\mathbf{r}} = \mathbf{r} - \mathbf{r}_t  \label{r_trans} \\
    \mathbf{r}' &\rightarrow \tilde{\mathbf{r}}' = \mathbf{r}' - \mathbf{r}_t \label{r'_trans}
\end{align}
where tildes denote translated coordinates, then we have
\begin{align}
    \tilde{U}(\tilde{\mathbf{r}}) &=
    \frac{I}{4\pi} \int_0^\infty \int_\mathcal{C} \frac{(\mathbf{r}-\mathbf{r}')\cross \mathbf{e}_{\tilde{r}} \cdot d\mathbf{l}'}{\abs{\mathbf{r}+t\mathbf{e}_{\tilde{r}} - \mathbf{r}'}^3} \ dt \label{eq:basic_U_explicit} \\
    &= \frac{1}{\mu_0} \int_0^\infty \mathbf{B} (\mathbf{r} + t \mathbf{e}_{\tilde{r}}) \cdot \mathbf{e}_{\tilde{r}} \ dt \label{path_int_ref} \\ 
    &= U(\mathbf{r}) \label{U_equivalence}
\end{align}
where the last equality holds by the fundamental theorem of calculus since $\mathbf{B}$ due to a closed current loop is conservative. Since the scalar potentials are equivalent, \eqref{BgradU} indicates that the magnetic fields corresponding to the original and translated coordinate systems are necessarily equal, i.e. $\tilde{\mathbf{B}}(\tilde{\mathbf{r}}) = \mathbf{B}(\mathbf{r})$. Therefore, under the electro-quasi-static assumption in a source-free region, there is translational symmetry for the magnetic potentials and magnetic fields produced by closed current loops. This is an obvious result, but it does not extend to the consideration of open segments that make up the closed loop, as shown next. For clarity, we will denote the magnetic scalar potential and magnetic field for open segments with lowercase letters $u(\mathbf{r})$ and $\mathbf{b}(\mathbf{r})$ respectively.

\subsection{Translational non-invariance of \texorpdfstring{$u(\mathbf{r})$}{Ur} and \texorpdfstring{$\mathbf{b}(\mathbf{r})$}{Br} for open current segments} \label{sec:open_segment}

Now, instead of a closed loop $\mathcal{C}$ we consider an open segment $\mathcal{D}$; the integral bounds of \eqref{eq:basic_U_explicit} become $\mathcal{D}$. Since \eqref{ampere} holds only for closed loops and does not necessarily hold true for open current segments, the magnetic field in this case is not necessarily conservative. Therefore, \eqref{U_equivalence} is not generally true for $u$, and the magnetic scalar potential is hence not always invariant under translations. I.e., $\tilde{u}(\tilde{\mathbf{r}})$ is not necessarily equal to $u(\mathbf{r})$ for open current segments.

We may explicitly write this result out as follows. Let use define 
\begin{align}
    \bar{F}(\mathbf{r},\mathbf{r}',\mathbf{r}_t) \equiv a(\tilde{r} a - ra - \mathbf{a} \cdot \mathbf{r}_t).
\end{align}
Under a coordinate translation by $\mathbf{r}_t$, we notice that
\begin{align}
    \tilde{F}(\tilde{\mathbf{r}},\tilde{\mathbf{r}}') &= F(\mathbf{r},\mathbf{r}') + \bar{F}(\mathbf{r},\mathbf{r}',\mathbf{r}_t) , \\
    d\tilde{\mathbf{L}}' &= -I(\tilde{\mathbf{r}}' \times d\mathbf{l}') \nonumber \\
    &= d\mathbf{L}' + I\mathbf{r}_t \times d\mathbf{l}',
\end{align}
and thus the translated version of \eqref{eq:novel_U} becomes
\begin{align}
    \tilde{u}(\tilde{\mathbf{r}}) &= -\frac{1}{4\pi}\int_\mathcal{D} \frac{d\tilde{\mathbf{L}}'\cdot \tilde{\mathbf{r}}}{\tilde{F}(\tilde{\mathbf{r}},\tilde{\mathbf{r}}')} \\
    &= -\frac{1}{4\pi}\int_\mathcal{D} \bigg[ \left(d\mathbf{L}' \cdot \mathbf{r} - (d\mathbf{L}' + I \mathbf{r} \times d\mathbf{l}') \cdot \mathbf{r}_t\right) \nonumber \\
    &\qquad \cdot \left(\frac{1}{F(\mathbf{r},\mathbf{r}')} + \frac{\bar{F}(\mathbf{r},\mathbf{r}',\mathbf{r}_t)}{F(\mathbf{r},\mathbf{r}')\tilde{F}(\tilde{\mathbf{r}},\tilde{\mathbf{r}}')} \right) \bigg] \\ 
    &= u(\mathbf{r}) + \bar{u}(\mathbf{r},\mathbf{r}_t), \label{U_tilde}
\end{align}
where
\begin{align}
    \bar{u}(\mathbf{r},\mathbf{r}_t)
    &=\frac{1}{4\pi}\int_\mathcal{D} \left[ \frac{d\mathbf{L}' \cdot \left(\bar{F}(\mathbf{r},\mathbf{r}',\mathbf{r}_t) \mathbf{r} + F(\mathbf{r},\mathbf{r}') \mathbf{r}_t\right)}{F(\mathbf{r},\mathbf{r}')\tilde{F}(\tilde{\mathbf{r}},\tilde{\mathbf{r}}')} \right. \nonumber \\
    & \hspace{1.75cm} + \left. \frac{I \mathbf{r} \times d\mathbf{l}' \cdot \mathbf{r}_t}{ \tilde{F}(\tilde{\mathbf{r}},\tilde{\mathbf{r}}')} \right]\label{Ubar}
\end{align}
is not necessarily zero. As such, open current segment contributions toward the magnetic scalar potential are not necessarily unique under coordinate translations. As a simple check, we see that when there is no coordinate translation, i.e. when $\mathbf{r}_t = \mathbf{0}$, then $\bar{F} = 0$, and hence $\bar{u} = 0$. This means that the scalar potential reduces to $\tilde{u} (\tilde{\mathbf{r}}) = u(\mathbf{r})$ as expected.

Now, we show that open segments of current loops do not have a unique contribution to $\tilde{\mathbf{b}}$ either. A simple rearrangement of \eqref{U_tilde} shows us that $u = \tilde{u} - \bar{u}$. Clearly then, $\grad_\mathbf{r} \bar{u}(\mathbf{r},\mathbf{r}_t) \neq \mathbf{0}$, since
\begin{align}
    \mathbf{b}(\mathbf{r}) = - \mu_0\grad_\mathbf{r} u(\mathbf{r}) \neq - \mu_0 \grad_\mathbf{r} \tilde{u}(\tilde{\mathbf{r}}).
\end{align}
This indicates that a translation by $-\mathbf{r}_t$ results in a difference of the magnetic field by $-\grad_\mathbf{r} \bar{u}(\tilde{\mathbf{r}},\mathbf{r}_t)$. 

As mentioned, the $d\mathbf{L}'$ in \eqref{eq:novel_U} and \eqref{eq:novel_B} indicate that radial current segments (when $d\mathbf{l}'$ is parallel to $\mathbf{r}'$) have zero $u(\mathbf{r})$ and $\mathbf{b}(\mathbf{r})$ contributions. It is clear that in the translated coordinate system where $\mathbf{r}' \rightarrow \tilde{\mathbf{r}}'$, this statement still holds true. Whenever $d\mathbf{l}'$ is parallel to $\tilde{\mathbf{r}}'$, the corresponding $\tilde{u}(\tilde{\mathbf{r}})$ and $\tilde{\mathbf{b}}(\tilde{\mathbf{r}})$ are zero.

\begin{figure}[htpb]
\centering
    \includegraphics[width=0.48\textwidth]{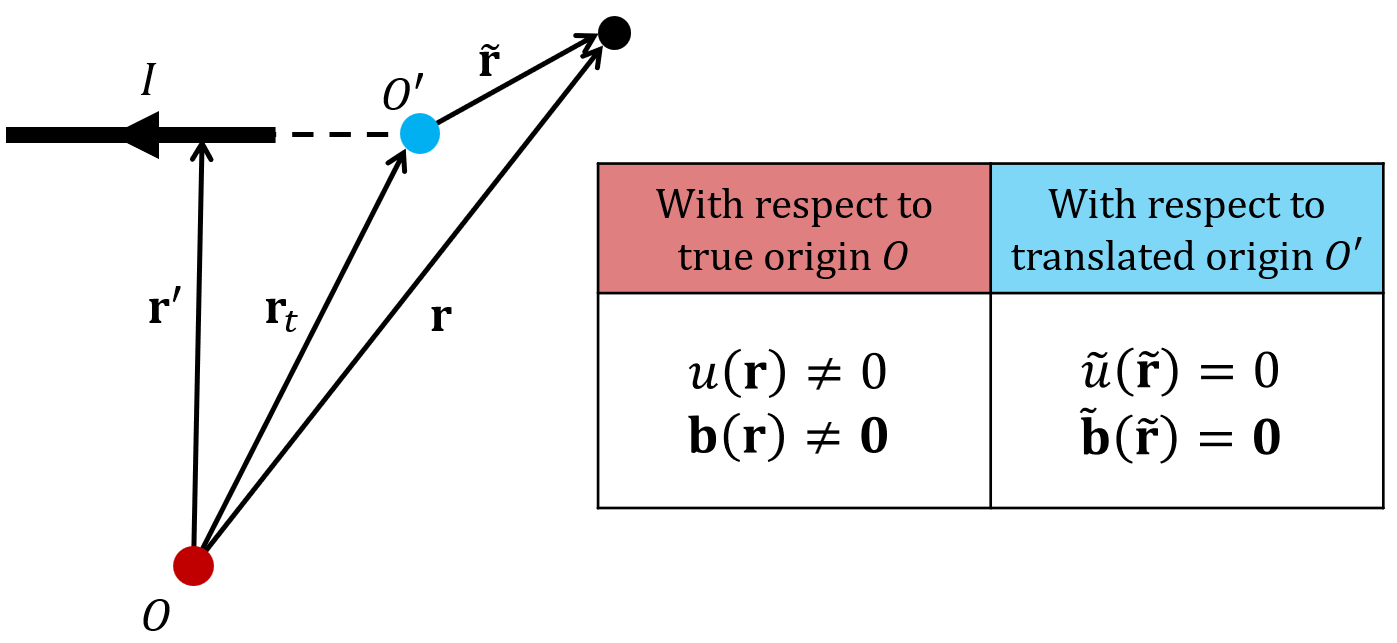}
    \caption{Illustration on how an open current segment has a non-unique magnetic scalar potential and magnetic field when different origins are chosen. In the coordinate system with origin $O$, the current segment is non-radial and hence not magnetically silent. However, with respect to the translated origin $O'$, the current segment becomes radial and is magnetically silent because $d\mathbf{L}'=0$ when $\mathbf{r}'$ is parallel to $d\mathbf{l}'$.}
    \label{fig:1}
\end{figure}

Figure~\ref{fig:1} summarizes the main results for this section (Section \ref{sec:open_segment}).

As an aside, in Appendix~\ref{sec:appendix_Efield}, we show that in the electric forward problem, the electric scalar potential and electric field are always translationally invariant under the magneto-quasi-static assumption for any segments of charge constituting the total charge configuration. 

\section{Significance of non-unique current segment contributions: MEG as an example case} \label{sec:meg}

In MEG, sensors are located outside of the head which is assumed to be a source-less region. Moreover, the electro- and magneto-quasi-static approximations are valid \cite{hamalainen1993magnetoencephalography}, thus the formulations so far (with the electro-quasi-static approximation) hold. In this section, we discuss relevant concepts in MEG, and extend on them with results presented in Section~\ref{sec:theory}.

\subsection{The primary and volume current}

In MEG, it is customary to express the total current density $\mathbf{J} (\mathbf{r}')$ as the sum of a primary current component $\mathbf{J}_{pri} (\mathbf{r}')$ and a volume current component $\mathbf{J}_{vol} (\mathbf{r}')$,
\begin{align} \label{J_tot}
    \mathbf{J} (\mathbf{r}') = \mathbf{J}_{pri} (\mathbf{r}') + \mathbf{J}_{vol} (\mathbf{r}').
\end{align}
The primary current corresponds to active brain sources, whereas the volume current corresponds to the passive return currents that complete the current loop \cite{hamalainen1993magnetoencephalography,ilmoniemi2019brain}. The current density is defined within the head, which is commonly modeled as a multi-layered conducting medium with closed and bounded surfaces \cite{hamalainen1989realistic,mosher1999eeg,de2012eeg,vorwerk2014guideline,stenroos2014comparison,nummenmaa2013comparison}.

The primary current density $\mathbf{J}_{pri} (\mathbf{r}')$ is often modeled as one or more focal sources, with each focal source represented by either a short and straight line segment, or an electric current dipole. If we assume $N$ dipolar sources, each of these focal dipolar current source may be written as $\mathbf{D}_i \delta (\mathbf{r}' - \mathbf{r}'_{D,i})$, where $\mathbf{D}_i$ and $\mathbf{r}'_{D,i}$ are the moment and location of the $i$\textsuperscript{th} current dipole respectively, $i = 1, \dots, N$. The primary current density in this case will thus be 
\begin{align} \label{Jp}
    \mathbf{J}_{pri} (\mathbf{r}') = \sum_{i=1}^N \mathbf{D}_i \delta (\mathbf{r}' - \mathbf{r}'_{D,i})
\end{align}
where $\delta(\mathbf{x})$ refers to the Dirac delta function. 

The volume current can be written as $\mathbf{J}_{vol} (\mathbf{r}') = \sigma (\mathbf{r}') \mathbf{E} ( \mathbf{r}')$, where $\sigma(\mathbf{r}')$ is the conductivity at $\mathbf{r}'$. In the magneto-quasi-static approximation, Faraday's equation reads $\grad \times \mathbf{E} (\mathbf{r}) = \mathbf{0}$ and thus we can write $\mathbf{E}$ as the gradient of an electric scalar potential $V$, $\mathbf{E} (\mathbf{r}) = - \grad V (\mathbf{r})$. Hence, the volume current component may be written as $\mathbf{J}_{vol} (\mathbf{r}') = -\sigma (\mathbf{r}') \grad V(\mathbf{r}')$.

\subsection{Geselowitz's formula}

Now, assume that the current density \eqref{J_tot} is within a bounded conductor $G$ with piecewise conductivities $\sigma_j$ in $n$ regions, $j = 1, \dots, n$, and each region bounded by surface $S_j$. From the decomposition of the total current density into primary and volume portions, the magnetic field may be expressed as
\begin{align} \label{geselowitz}
    \mathbf{B} (\mathbf{r}) = \mathbf{b}_{pri} (\mathbf{r}) + \mathbf{b}_{vol} (\mathbf{r})
\end{align}
where 
\begin{widetext}
\begin{align}
    \mathbf{b}_{pri} (\mathbf{r}) &= \frac{\mu_0}{4 \pi} \int_G \mathbf{J}_{pri} \left(\mathbf{r}'\right) \times \frac{\mathbf{r} - \mathbf{r}'}{\abs{\mathbf{r} - \mathbf{r}'}^3} \ dv' \label{B_pri_g} \\
     \mathbf{b}_{vol} (\mathbf{r}) &= - \frac{\mu_0}{4 \pi} \sum_{j=1}^n \left(\sigma_j' - \sigma_j''\right) \int_{S_j} V\left(\mathbf{r}'\right) \mathbf{n}' \left(\mathbf{r}'\right) \times \frac{\mathbf{r} - \mathbf{r}'}{\abs{\mathbf{r} - \mathbf{r}'}^3} \ dS_j', \label{B_vol_g}
\end{align}
\end{widetext}
and $\mathbf{n}'(\mathbf{r}')$ is the outward unit normal of surface $S_j$, $\sigma_j'$ and $\sigma_j''$ are the inner and outer conductivities with respect to $S_j$, and $V(\mathbf{r}')$ is the electric scalar potential on $S_j$.

Equation \eqref{geselowitz}, with its terms defined by \eqref{B_pri_g} and \eqref{B_vol_g}, is known as \textit{Geselowitz's formula} \cite{geselowitz1967bioelectric}. In this formula, $\mathbf{b}_{pri}$ \eqref{B_pri_g} corresponds to the primary current contribution to the magnetic field, whereas $\mathbf{b}_{vol}$ \eqref{B_vol_g} corresponds to the volume current contribution. Geselowitz's formula can be derived by substituting the total current expression \eqref{J_tot} into Biot-Savart law \eqref{biot}, then using vector identities including Stoke's theorem to obtain its final form. Note that Geselowitz's formula implies that volume currents can be equivalently expressed as surface current densities on $S_j$ with orientations $\mathbf{n}'(\mathbf{r}')$.

If we assume $N$ dipole sources as in \eqref{Jp}, then the primary current contribution of Geselowitz's formula \eqref{B_pri_g} collapses due to the Dirac delta function:
\begin{align} \label{Bpri_dip}
    \mathbf{b}_{pri} (\mathbf{r}) = \frac{\mu_0}{4 \pi} \sum_{i=1}^N \mathbf{D}_i \times \frac{\mathbf{r} - \mathbf{r}'_{D,i}}{\abs{\mathbf{r} - \mathbf{r}'_{D,i}}^3}.
\end{align}
We will assume $N = 1$ subsequently and drop the dipole-specific subscripts in order to simplify our investigation on the impacts of the translationally non-invariant current segment contributions $\mathbf{b}_{pri}$ and $\mathbf{b}_{vol}$.

\subsection{The magneto- and electro-quasi-static form of Geselowitz's formula (Unified Geselowitz-Sarvas formula)} \label{sec:vol_to_line}

Geselowitz's formula assumes the magneto-quasi-static approximation only. Here, we show the form of it that satisfies the electro-quastatic approximation as well. This unifies Geselowitz's formula and Sarvas' formula (without the assumption of a dipolar current source), since Sarvas' formula assumes the electro-quasi-static approximation only. By assuming that \eqref{geselowitz} can be written in the form of \eqref{BgradU} and following \eqref{U_integral}, we see that the magnetic scalar potential for the primary and volume contributions are
\begin{widetext}
\begin{align}
    u_{pri}(\mathbf{r}) &= -\frac{1}{4 \pi} \int_G \frac{\mathbf{J}_{pri}(\mathbf{r}') \times \mathbf{r}' \cdot \mathbf{r}}{F(\mathbf{r},\mathbf{r}')} \ dv' \label{U_pri} \\
    u_{vol} (\mathbf{r}) &= \frac{1}{4 \pi}  \sum_{j=1}^n \left(\sigma_j' - \sigma_j''\right) \int_{S_j} V\left(\mathbf{r}'\right) \frac{ \mathbf{n}'\left(\mathbf{r}'\right) \times \mathbf{r}' \cdot  \mathbf{r}}{F \left(\mathbf{r},\mathbf{r}'\right)} \ dS_j' \label{U_vol}
\end{align}
\end{widetext}
and the magnetic field contributions are thus
\begin{widetext}
\begin{align}
    \mathbf{b}_{pri}(\mathbf{r}) &= \frac{\mu_0}{4 \pi} \int_G \left( \frac{F(\mathbf{r},\mathbf{r}') \mathbf{J}_{pri} \times \mathbf{r}' - \mathbf{J}_{pri} \times \mathbf{r}' \cdot \mathbf{r} \grad F(\mathbf{r},\mathbf{r}')}{ F^2(\mathbf{r},\mathbf{r}')} \right) \ dv' \label{B_pri} \\
    \mathbf{b}_{vol}(\mathbf{r}) &= - \frac{\mu_0}{4 \pi} \sum_{j=1}^n \left(\sigma_j' - \sigma_j''\right) \int_{S_j} V\left(\mathbf{r}'\right) \frac{F(\mathbf{r},\mathbf{r}') \mathbf{n}'\left(\mathbf{r}'\right) \times \mathbf{r}' - \mathbf{n}'\left(\mathbf{r}'\right) \times \mathbf{r}' \cdot \mathbf{r} \grad F(\mathbf{r},\mathbf{r}')}{F^2(\mathbf{r},\mathbf{r}')} \ dS_j'. \label{B_vol}
\end{align}
\end{widetext}
As a check, we verify that when we have radially-oriented segments, i.e. when $\mathbf{J}_{pri}$ is parallel to $\mathbf{r}'$ in the primary current case or when $\mathbf{n}'(\mathbf{r}')$ is parallel to $\mathbf{r}'$ in the volume current case, $\mathbf{b}_{pri} = \mathbf{0}$ and $\mathbf{b}_{vol} = \mathbf{0}$, respectively, as expected from Section~\ref{sec:forward}.

\subsection{The current triangle formulation for spherical head models} \label{sec:current_triangle}

So far, we have stated that a focal primary current source may be modeled as a finite-length current segment of current $I$. The start and end of the primary current segment represent a sink and a source, respectively \cite{ilmoniemi1985forward,ilmoniemi2009triangle}. Although our formulations in Section~\ref{sec:theory} hold for current loops, they still hold in the context of MEG where volume currents are of interest. From Kircchoff's junction rule, we know that the volume current can be written as a finite sum of line currents flowing from the source to the sink defined by the primary current (equation \eqref{eq:uvol_line_sum}). Thus, the magnetic contribution from the volume current can in theory be expressed as a superposition of the contributions from each of the current loops. 

As an aside, we attempted to further generalize to show that there exists an equivalent line current that produces the magnetic field of any volume current. In Appendix \ref{sec:volume_current_line}, we show that such an equivalent line current exists for each field point; however, it appears that it may be non trivial to show if there exists a common equivalent line current that holds for all field points.

Here, we show that in combining our framework with the case of a spherical head model, the current density can always be written as a triangle loop with one vertex coinciding with the center of the conducting sphere. This is in agreement with the results presented in \cite{ilmoniemi1985forward,ilmoniemi2009triangle}. 

Consider a spherical conductor model with origin at the center of the sphere, i.e.
\begin{align} \label{normal_def}
    \mathbf{n}'(\mathbf{r}') = \frac{\mathbf{r}'}{r'}.
\end{align}
The volume contribution of Geselowitz's formula $\mathbf{b}_{vol}(\mathbf{r})$ \eqref{B_vol} vanishes due to the scalar triple product. Hence, for a spherical conductor head model, the external magnetic field has no volume current contribution if we take the origin to be at the center of the sphere, i.e. $\mathbf{B}(\mathbf{r}) = \mathbf{b}_{pri}(\mathbf{r})$ everywhere outside the conductor ($\mathbf{b}_{vol}(\mathbf{r}) = \mathbf{0}$). This applies to all three components of the magnetic field.

We also know from previous sections that radial current segments give zero magnetic field contributions in our formulation. This means that the volume current can always be modeled as radial segments that connect from the center of the sphere to either ends of the primary current line segment, forming a current triangle.

If we shift the origin away from the center of the sphere, the statement that closed loops give translationally invariant contributions implies that the equivalent current triangle model is still valid. The fact that the radial volume current segments now become non-radial, leading to $\tilde{\mathbf{b}}_{vol} \neq \mathbf{0}$, is compensated by the primary current contribution changing due to a change in orientation. In other words, $\mathbf{b}_{pri} \neq \tilde{\mathbf{b}}_{pri} = \mathbf{b}_{pri} - \tilde{\mathbf{b}}_{vol}$ to ensure $\mathbf{B}=\mathbf{b}_{pri}+\mathbf{b}_{vol} = \tilde{\mathbf{b}}_{pri} + \tilde{\mathbf{b}}_{vol}$. The primary and volume current contributions in this new coordinate system can be obtained explicitly by substituting \eqref{r_trans} and \eqref{r'_trans} into \eqref{B_pri} and \eqref{B_vol} respectively.

If the conductor is not spherically symmetric, $\tilde{\mathbf{b}}_{vol}(\tilde{\mathbf{r}})\neq \mathbf{0}$ for all choices of origin, since there is no origin where all volume current can be represented with radial segments. 
 
This result also tells us that in the case of a spherical head geometry, the formulations from Section~\ref{sec:theory} hold, with $d\mathbf{l}' = d\mathbf{l}'_{pri} + d\mathbf{l}'_{vol}$. We may also easily see that \eqref{B_pri} is equivalent to \eqref{eq:novel_B} and \eqref{U_pri} is equivalent to \eqref{eq:novel_U}, since we can now write $\mathbf{J} dv' = \abs{\mathbf{J}} d\mathbf{l}' = I d\mathbf{l}'$. 

We note in Appendix~\ref{sec:appendix_greens_thm} that in the case where the closed loop is flat and the origin is in the plane of the loop, equations \eqref{eq:novel_U} and \eqref{eq:novel_B} can be written as a line integral with outward pointing normal. This is analogous to the reduction of the volume current contribution from a volume integral to a surface integral in Geselowitz's formula in 2D.

\subsection{Sarvas' formula}

In addition to assuming a spherical head model with a current triangle formulation, let us now assume that the primary current segment is infinitesimally short, i.e., a dipolar primary current source $\mathbf{D} \delta (\mathbf{r}' - \mathbf{r}'_{D})$. If we assume the origin to be at the center of the sphere, then the radial volume currents have zero contribution, and hence the total magnetic field outside the conductor is given by
\begin{align}
    \mathbf{B}(\mathbf{r}) &= \mathbf{b}_{pri} \nonumber \\
    = \frac{\mu_0 }{4\pi} &\left[  \frac{F(\mathbf{r},\mathbf{r}'_D) \mathbf{D} \times \mathbf{r}'_D - \mathbf{D} \times \mathbf{r}'_D \cdot \mathbf{r} \grad F(\mathbf{r},\mathbf{r}'_D)}{F^2(\mathbf{r},\mathbf{r}'_D)} \right], \label{eq:B_sarvas}
\end{align}
which may be obtained from either equivalent formulas \eqref{eq:novel_B} or \eqref{B_pri}. This is precisely \textit{Sarvas' formula} \cite{sarvas1987basic}.

\section{Verification of our approach using simulations} \label{sec: verifications}

In this section, we first briefly introduce ECD fits and their validity, then show the effect of origin shifts on the localization accuracy of a simple MEG ECD fit.

\subsection{ECD inverse models and their validity} \label{sec:ECD}

We have shown that individual segments of a current loop have non-unique contributions to the magnetic field under coordinate translations despite the fact that the contribution of the loop as a whole is translation-invariant. We now show the effects of this result on ECD inverse modeling for a spherical head model.

In general, ECD fits attempt to fit the signal produced by one or more electric dipole sources (calculated using Sarvas' formula \eqref{eq:B_sarvas}) to a reference signal. This is done by finding ideal locations for the dipoles so that the error between the dipoles' signal and reference signal is minimized. Its performance may depend on many factor, e.g., a sufficiently accurate initial guess, the objective function choice, and the stability of the solution. 

Two methods in which the ideal dipole locations may be found are by directly varying their locations in an iterative manner to minimize the signal error, or by setting up a grid of plausible dipole locations, then exhaustively finding which location produces the lowest signal error. The former iterative method has the advantage of being more computationally efficient, however it runs the risk of converging to a local minima that is not the desired solution. The latter circumvents this issue, but may not accurately find the global solution if the grid is coarse; having a fine grid also tends to be computationally slow \cite{cover2007fitting}. In our case, we employ the latter exhaustive search which will be detailed in Section~\ref{ecd_procedure} later on. This is because we are interested in the behavior of fits and the global solution, rather than absolute accuracy.

We investigated the validity of ECD fits by attempting to show if there exists an equivalent current dipole that produces the identical magnetic field of a straight line current. In Appendix \ref{sec:mean_value_theorem}, we show that for some for some field point, there always exists an equivalent current dipole for a straight current segment that lies on the segment itself and has a parallel orientation. However, similar to the situation of finding an equivalent line current for volume currents as discussed in Section~\ref{sec:current_triangle} and Appendix \ref{sec:volume_current_line}, it appears that it may be non-trivial to show if there exists an equivalent current dipole that holds for all field points.

Since it is uncertain if there exists common equivalent current dipole for all field points, as well as the fact that the mean value theorem only guarantees the existence of a solution along the line, ECD fits may thus be informative only for short line current segments such as our primary current choice in Section~\ref{ecd_procedure}. A short current segment restricts the position of the equivalent current dipole to have a small variation.

From our assessment, ECD fits are thus valid as an inverse source modeling option when the current triangle formulation is used, with the origin taken to be at the center of spherical head model and coinciding with a vertex of the triangle. Ideally, the edges are short as well. The radially-oriented volume currents will be magnetically silent, and hence the effective source current model is the primary current straight line segment. If the origin coincides with any one of the triangle's other vertices, then the opposite edge to this new origin, which was once radial but is now the only non-radial segment, now contributes entirely to the total magnetic field. An ECD fit is thus also valid in the sense that a dipole that represents the volume currents may be found, although it does not lie on the primary current edge that we seek. See Sections~\ref{ecd_results} and \ref{sec:discussion} for further results and details.

\subsection{Illustration of non-unique \texorpdfstring{$\tilde{u}(\tilde{\mathbf{r}})$}{Ur} contributions by current line segments} \label{sec:nonunique_U}

\subsubsection{Set-up and procedure}

We now illustrate the non-invariant magnetic scalar potential contribution $\tilde{u} (\tilde{\mathbf{r}})$ by open current line segments in translated coordinate systems, as discussed in Section~\ref{sec:open_segment}.

A triangular current loop was considered, and we have denoted each edge's contribution as $\tilde{u}_i (\tilde{\mathbf{r}})$,  $i = 1,2,3$. The triangle loop was defined by specifying three random vertices of up to length 5~cm from $(0,0,0)$~cm. We then considered 12 origins to evaluate the contributions of each edge from: 9 origins (labeled as origin indices 1 to 9) are spaced evenly on the triangle loop, and 3 origins (indices 10 to 12) of up to 5~cm from $(0,0,0)$~cm that do not lie on the triangle were randomly picked. The field point was randomly picked to be 12~cm away from $(0,0,0)$~cm, and the magnetic scalar potential was evaluated with \eqref{eq:novel_U}.

\subsubsection{Results}

\begin{figure}[htpb]
    \centering
    \includegraphics[width=0.48\textwidth]{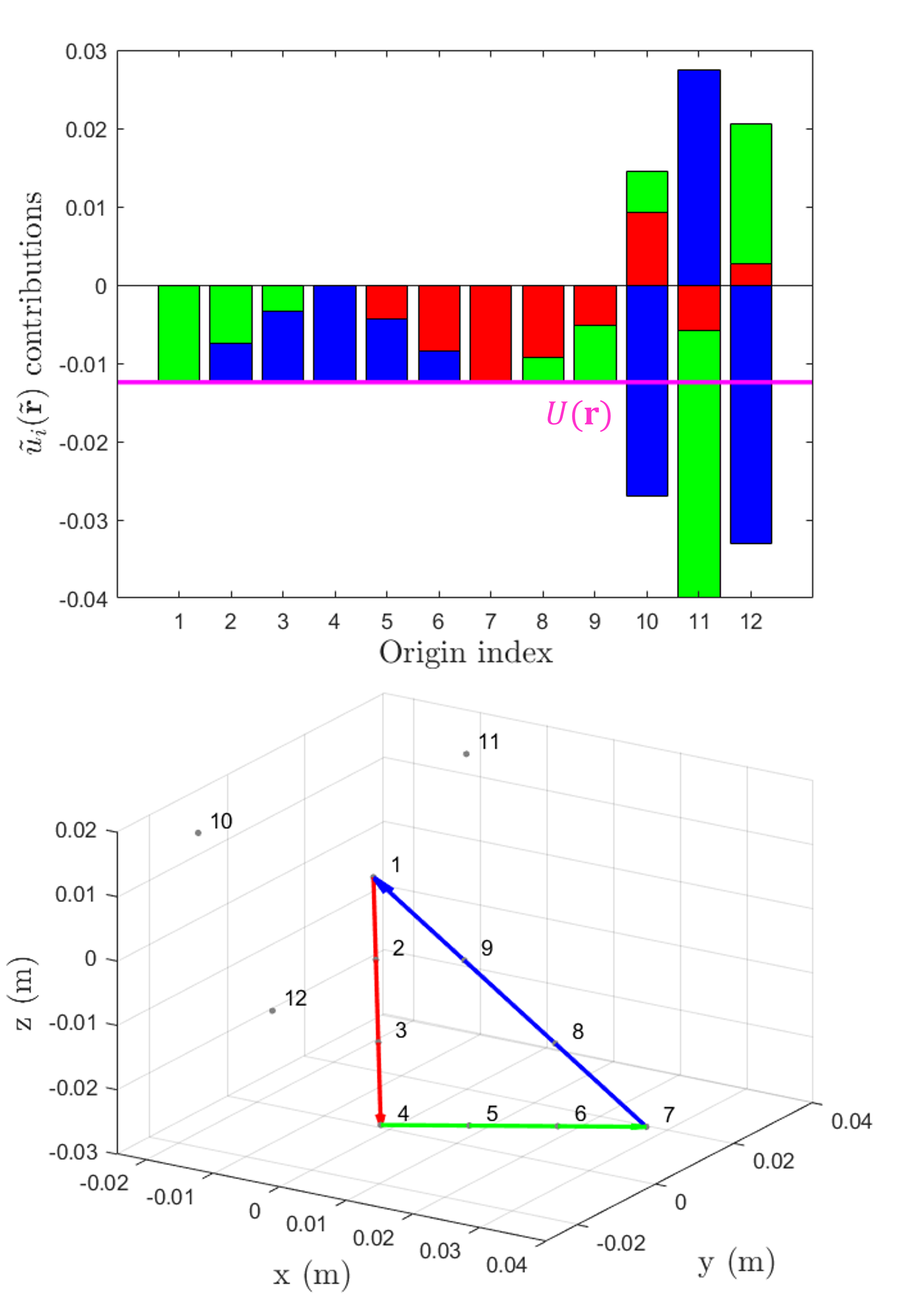}
    \caption{The top panel illustrates the varying $\tilde{u}_i (\tilde{\mathbf{r}})$ contribution by each line segment of the current triangle at various origins. Only non-radial current segments have non-zero contribution. Origin indices 1 to 9 shows origin shifts along the triangle, and indices 10 to 12 are three random off-triangle origins. The magenta line shows the translationally invariant $U(\mathbf{r})$ of the closed loop. The bottom panel shows the 12 origins considered. Each edge and their corresponding $\tilde{u}_i (\tilde{\mathbf{r}})$ contributions have matching colors.}
    \label{fig:2}
\end{figure}

The physical set-up of the triangle and origin points are shown in the bottom subplot of Figure~\ref{fig:2}, and the contributions from each edge when evaluated from each of the 12 origins using are illustrated in the top subplot of Figure~\ref{fig:2}.

We notice that indeed, $\tilde{u}_i (\tilde{\mathbf{r}})$ is not invariant for each edge under coordinate translation. However, when all three components are added to form a closed loop contribution (magenta line in Figure~\ref{fig:2}), this quantity corresponding to the total scalar potential is invariant under coordinate translation as expected from Section~\ref{sec:closed_loop}. Moreover, when the origins lie at the vertices or along an edge, we see that only the non-radial edges relative to that origin had a non-zero scalar potential contribution as expected from \eqref{eq:novel_U}.

\subsection{Explicit verification of ECD fit validity} \label{sec:ecd_verification}

\subsubsection{Set-up and procedure}

Here, we perform a simple verification of our claim in Section~\ref{sec:ECD} and Appendix~\ref{sec:mean_value_theorem} that for a triangular current loop with origin at any of its vertices, there always exists a dipole that lies in the opposite edge that exactly reproduces the magnetic field of that edge when evaluated at some fixed field point.

We first specified an arbitrary current triangle with each vertex coordinate up to 5~cm away from the $(0,0,0)$~cm origin and current strength $I = 1$~A. In reality, physiological dipole moments are best measured in units of nAm, but we used this arbitrary current strength value as it is an irrelevant parameter with respect to our dipole fitting procedure. Each edge was discretized into 50 points, and each point was taken to be an origin. The signal due to a dipole located at every other point was calculated, with orientation parallel to the edge it lies on and strength $I$ multiplied by that edge's length. Let us denote each of these signals to be $\bm{\phi}_{D}$.

The forward magnetic field due to the entire triangle used to calculate the reference flux signal was obtained by performing the Biot-Savart line integral over the closed loop. For convenience, the calculation was done with respect to the $(0,0,0)$~cm origin; we know from Section~\ref{sec:closed_loop} that this calculation over the triangle loop will yield translationally invariant results. We will also refer to this signal as $\bm{\phi}_{ref}$ from hereon.

The MEG sensor array considered was the standard placement of the 102 magnetometers of the Elekta Neuromag TRIUX system. Each magnetometer is square with side length 2.1~cm, and the magnetic flux or signal for each sensor was calculated via a 9-point cubature sampling approximation of the magnetic field over the sensor area \cite{abramowitz1988handbook}. The sensor array is illustrated in the top subplot of Figure~\ref{fig:4}.

Let the error between $\bm{\phi}_{ref}$ and $\bm{\phi}_{D}$ be quantified by 
\begin{align} \label{objective_fn}
    \epsilon = \frac{\norm{\bm{\phi}_{ref} - \bm{\phi}_{D}}^2}{\norm{\bm{\phi}_{ref}}^2}
\end{align}
which ranges from $\epsilon = 0$ when $\bm{\phi}_{ref} = \bm{\phi}_{D}$ to $\epsilon = 1$ when $\bm{\phi}_{D} = \mathbf{0}$ or $\bm{\phi}_{D} = -2 \bm{\phi}_{ref}$. This means that we may define a goodness of fit (GOF),
\begin{align} \label{GOF}
    \text{GOF} = 1 - \epsilon,
\end{align}
with $\text{GOF} = 0$ indicating the worst possible agreement between $\bm{\phi}_{ref}$ and $\bm{\phi}_{D}$, and $\text{GOF} = 1$ indicating perfect agreement.

\subsubsection{Results}

\begin{figure}[htpb] 
\centering
    \includegraphics[width=0.48\textwidth]{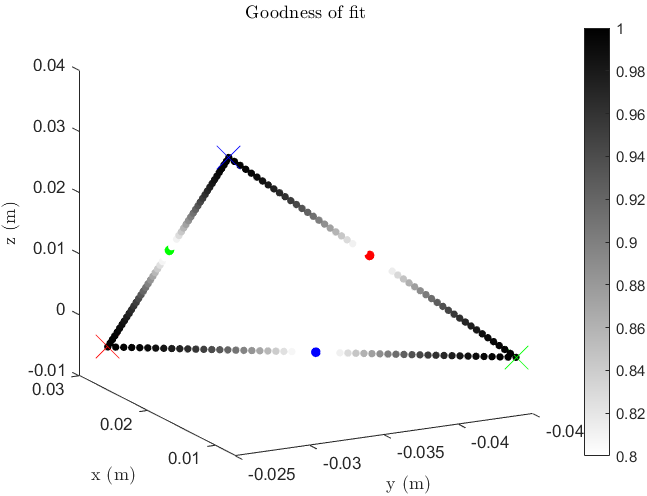}
    \caption{Shows the maximum ECD GOF attainable at each origin point that lies along an arbitrary current triangle, when the dipoles are constrained to be parallel along the triangle edges. In the cases where the origin coincides with a triangle vertex, we see that the dipole with a maximum GOF of approximately 1 lies on the opposite edge near the center. Each origin is marked by an ``x'', and the dipole location is marked by a solid circle with matching color. This is in agreement with the results in Section~\ref{sec:mean_value_theorem}.}
    \label{fig:3}
\end{figure}

Figure \ref{fig:3} shows the highest GOF value found at every origin for some dipole along the triangle. We see that indeed, when the origins are at the triangle vertices, $\text{GOF} \approx 1$. Each vertex origin is marked by a colored ``x'', and the dipole that yielded the highest GOF value is marked with a solid circle with the corresponding color. Again, these dipoles reside on the opposite edge as expected.

We also see that in the case where the origin does not coincide with any of the triangle's vertices but lie on the triangle, the highest attainable GOF for dipole sources parallel along the triangle edges is less than 1, reaching a minimum for origins near the center of each edge.

In Section~\ref{ecd_results} later on however, we observe via an unconstrained ECD fit simulation that there exist dipoles not necessarily lying on the triangle that yield a high GOF when the origin lies in the plane of the triangle. This is shown in the bottom left plot of Figure~\ref{fig:5}, and further discussed in Section~\ref{sec:discussion}.

\subsection{Unconstrained ECD fit procedure} \label{ecd_procedure}

\subsubsection{Set-up}

Now, we illustrate how the property of open current segments having translationally non-invariant contributions affects a simple unconstrained MEG ECD fit.

\begin{figure}[htpb] 
\centering
    \includegraphics[width=0.48\textwidth]{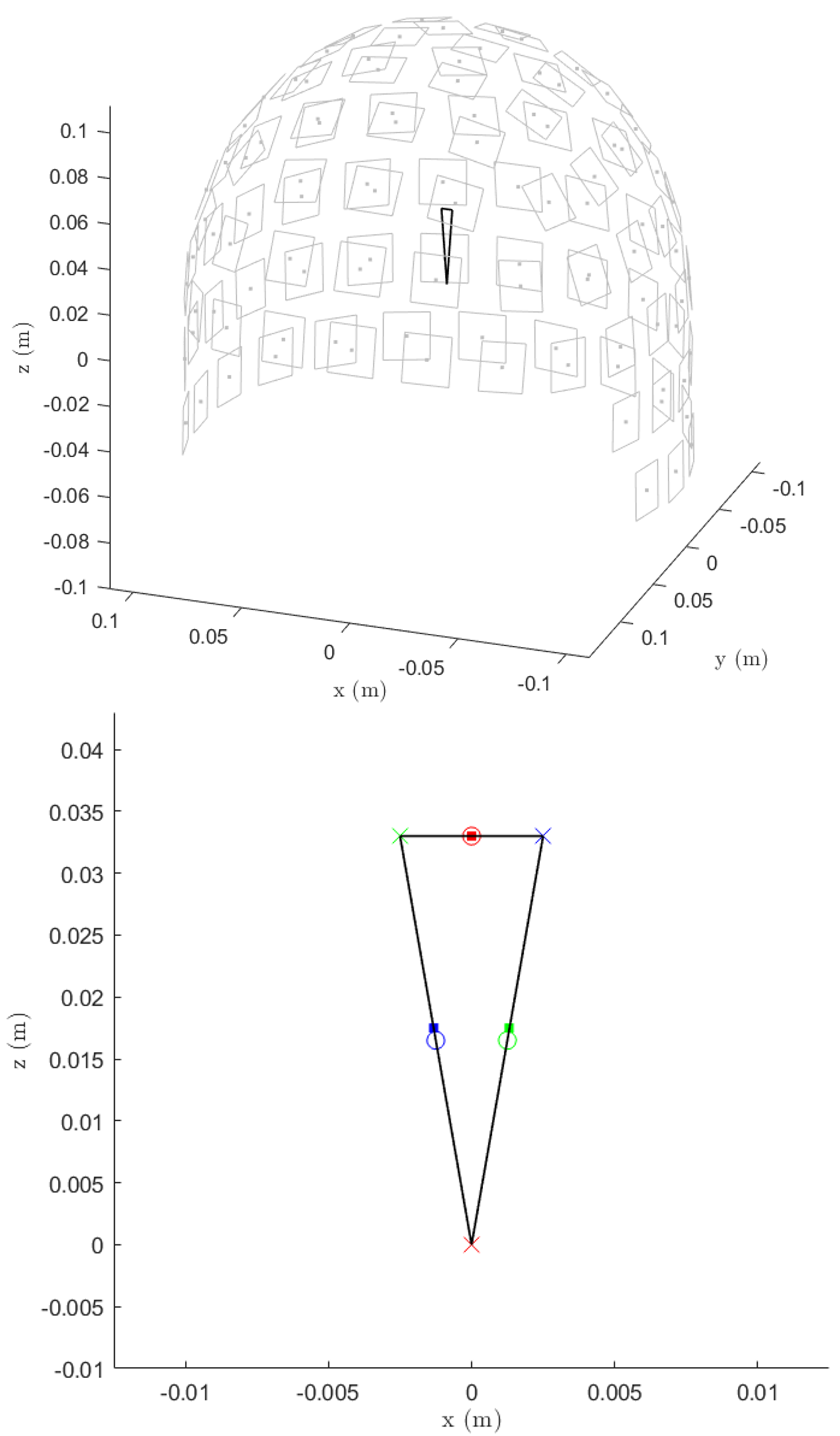}
    \caption{(Top) Illustrates the magnetometer sensor array set-up and the current triangle configuration for Section~\ref{ecd_procedure}. (Bottom) Unconstrained ECD fit results when the origin choice coincides with one of the triangle vertices. Each origin is marked by an ``x'', and the corresponding ECD fit result is marked by a solid square with matching color. The center of each triangle edge is marked by a hollow circle. This agrees with the constrained ECD result shown in Figure~\ref{fig:3}.}
    \label{fig:4}
\end{figure}

We again considered a spherical head model with a current triangle loop source. The three triangle vertices were defined to be at $(0,0,0)$~cm and $(\pm 0.25,0,3.3)$~cm, corresponding to the deep-source configuration of the Elekta Neuromag (Megin Oy, Helsinki, Finland) dry phantom. Conventionally, the short horizontal edge is interpreted to be primary current, with origin set at $(0,0,0)$~cm. This current triangle set-up is shown in the top subplot of Figure~\ref{fig:4}, relative to the sensor array.

Let us denote the ``reference origin'' as the center of the sphere, i.e., $(0,0,0)$~cm, coinciding with one of the triangle vertices. The current strength was arbitrarily set to 1~A; its value has no effect on the behavior of the results. The sensor array and method of calculating the forward signal was identical to what was used in the previous section (Section~\ref{sec:ecd_verification}).

Two planes of $20 \times 20$ grids of translated origins was defined: one in the $y=0$ ($xz)$ plane from $x \in [-3,3]$~cm and $z \in [-1,5]$~cm, and the other in the $z=0$ ($xy$) plane from $x$, $y \in [-3,3]$~cm. These planes are parallel and perpendicular to the plane defined by the current triangle, respectively. 

\subsubsection{Procedure}

As mentioned in Section~\ref{sec:ECD}, we opted for an exhaustive search for the optimal dipole location instead of an iterative search. In our context, this means that every grid point in both the $xz$ and $xy$ planes was set to be a (translated) origin, and for each of these origins, an optimal $xz$ grid point not equal to itself was identified as the ECD location with some orientation that minimizes the error between $\bm{\phi}_{ref}$ and $\bm{\phi}_{D}$ there.

Let us consider one particular origin and $xz$ grid point not equal to the origin. First, the corresponding topography $\bm{\gamma}_D$ was calculated, which contains the signal distribution of the dipole across all sensors. This topography is a matrix of size $102 \times 3$, with each column corresponding to the signal measured by the 102 magnetometer sensor array due to a unit dipole located at that grid point oriented in each of the 3 orthogonal coordinate directions. By this definition of $\bm{\gamma}_D$, if we define $\bm{\alpha}_D$ to be a $3 \times 1$ vector containing the dipole strength in each of the 3 orthogonal coordinate directions, the signal due to a dipole located at the grid point will be
\begin{align}
    \bm{\phi}_{D} = \bm{\gamma}_D \bm{\alpha}_D.
\end{align}
Note that $\bm{\alpha}_D$ is in theory unit-less; however, it has exactly the same values as the dipole moment. In the following, we will simply refer to $\bm{\alpha}_D$ as the dipole moment.

The dipole moment $\bm{\alpha}_D = \bm{\alpha}_{D,est}$ that minimizes the objective function $\epsilon$ \eqref{objective_fn} was then obtained,
\begin{align}
    \bm{\alpha}_{D,est} = \argmin_{\bm{\alpha}_D} \left( \frac{\norm{\bm{\phi}_{ref} - \bm{\gamma}_D \bm{\alpha}_D}^2}{\norm{\bm{\phi}_{ref}}^2} \right).
\end{align}
This was done using Matlab's fminsearch with an arbitrary initial guess of $\bm{\alpha}_D = (1,1,1)$; we found that our results obtained were not sensitive to the initial guess. Then, the GOF was found with \eqref{GOF},
\begin{align}
    \text{GOF} = 1 - \epsilon(\bm{\alpha}_{D,est}).
\end{align}

This process is repeated for all other grid points for this particular origin. The grid point and optimized dipole moment corresponding to the highest GOF was determined to be the ECD solution for this origin. Let us denote this highest GOF value as $\text{GOF}_{ECD}$, and the corresponding optimal dipole position and moment as $\mathbf{r}'_{ECD}$ and $\bm{\alpha}_{ECD}$ respectively.

Note that the immediate solution $\mathbf{r}_{ECD}'$ that an ECD fit provides is relative to the translated origin. In order to compare this with the expected true solution $\mathbf{r}_{D,true}'$ which is relative to the reference origin, one needs to translate the solution $\mathbf{r}_{ECD}'$ back into the coordinate system defined by the reference origin, or vice versa. For notational simplicity, all mentions of $\mathbf{r}_{ECD}'$ from now on refer to the ECD fit solution that has already been appropriately translated back into the reference coordinate system.

Finally, the localization error (LE) of the ECD solution was found via taking the Euclidean distance between $\mathbf{r}_{ECD}'$ and an expected ECD solution $\mathbf{r}_{D,true}'$,
\begin{align}
    \text{LE}_{ECD} = \norm{ \mathbf{r}_{ECD}' - \mathbf{r}_{D,true}'}.
\end{align}
The expected solution was set to be at the center of the short horizontal edge, $\mathbf{r}_{D,true}' = (0,0,3.3)$~cm. This is determined from our results as shown in Figure~\ref{fig:3}, which indicates that the exact ECD fit should lie approximately in the center of the edge.

The above procedure is repeated for all other origins. The final result is thus a $\text{GOF}_{ECD}$, $\text{LE}_{ECD}$, $\mathbf{r}_{ECD}'$, and $\bm{\alpha}_{ECD}'$, assigned to each origin.

Also, note that there are other objective function choices that one could minimize. For example, the subspace angle between $\bm{\phi}_{ref}$ and $\bm{\phi}_{D}$ can be iteratively minimized by varying the dipole position $\mathbf{r}_D$. We found that the results using this method managed to avoid converging to any potential local minima, and yielded similar results that are to follow.

\subsubsection{Results} \label{ecd_results}

\begin{figure}[htpb]
\centering
    \includegraphics[width=0.48\textwidth]{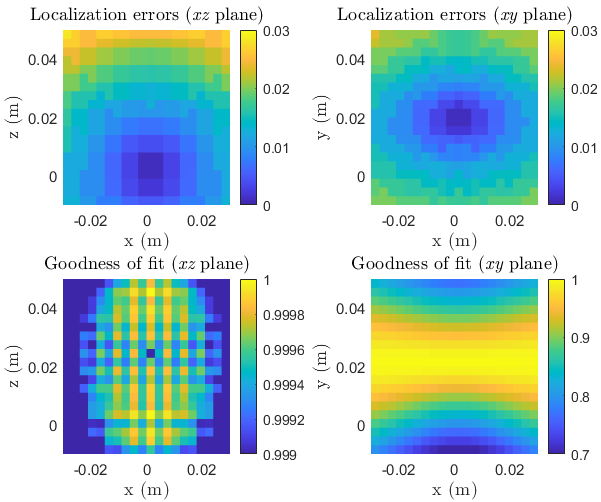}
    \caption{$\text{LE}_{ECD}$ and $\text{GOF}_{ECD}$ plots for the ECD fits at various origin choices. The current triangle was defined at $(0,0,0)$~cm and $(\pm 0.25,0,3.3)$~cm on the $xz$ plane. LE increases when the origin choice is further from the reference (0,0,0) cm origin, but GOF is high for all origins in the $xz$ plane, especially in areas near the triangle vertices. GOF falls off much quicker for origin translations in directions perpendicular to the triangle plane. LE is comparable for origin translations of the same amount in both the $xz$ plane and the $xy$ plane.} 
    \label{fig:5}
\end{figure}

The $\text{LE}_{ECD}$ and $\text{GOF}_{ECD}$ results of the ECD fit for each origin that lie on the $xz$ and $xy$ planes are shown in Figure~\ref{fig:5}.

The first column of Figure~\ref{fig:5} shows the $\text{LE}_{ECD}$ and $\text{GOF}_{ECD}$ values in the $xz$ plane. Despite the LE having a minimum only at the reference origin, the GOF has a maxima not only at this reference origin, but also the region around the other two triangle vertices. This may be because ECD fits attempt to find an electric dipole, or equivalently, a straight line segment as discussed in Section~\ref{sec:mean_value_theorem}, that produces a signal that most closely replicates the measured signal. If the origin coincides with any of the three triangle vertices, only the opposite edge is non-radial and thus has total signal contribution, resulting in the ECD fit being able to find a solution with high GOF that lies on the opposite edge.

The locations of the ECD fits when the origin coincides with the three triangle vertices are shown in the bottom subplot of Figure~\ref{fig:4}. Each origin is denoted by a colored ``x'', and the corresponding ECD fit is denoted by a solid square marker with the matching color. The hollow circles denote the center of each edge. We see that indeed, the ECD fits localized the dipole to the opposite edge as expected, which is consistent with the theory in Section~\ref{sec:mean_value_theorem} and its verification in Section~\ref{sec:ecd_verification}.

The second column of Figure~\ref{fig:5} shows the LE and GOF in the $xy$ plane. We see that in this plane that is perpendicular to the current triangle, the LE amounts are similar to the results in the parallel $xz$ plane. However, the GOF decays much more rapidly away from the reference origin. This may be explained by how a single dipole/current line segment cannot replicate the signal produced by more than one current segment when the origin does not coincide with one of the three triangle vertices and does not lie on the triangle's plane.

From the LE and GOF plots, we see that having a high ECD GOF is not an accurate indication of low LE. GOF is anisotropic, especially in the case where the origin is in the plane parallel to the current triangle, and is generally not an accurate indicator of LE in any way.

\subsubsection{Origins in the plane of the triangle} \label{sec:inverse_relation}

We note however that in the bottom left subplot of Figure~\ref{fig:5}, the GOF values are approximately 1 for all origins in the $xz$ plane. Upon further inspection of $\bm{\alpha}_{ECD}'$, we notice that the ECD orientations were all approximately in the $xz$ plane as well, implying that there is a near-perfect ECD fit in the plane of the triangle. 

\begin{figure}[htpb] 
\centering
    \includegraphics[width=0.48\textwidth]{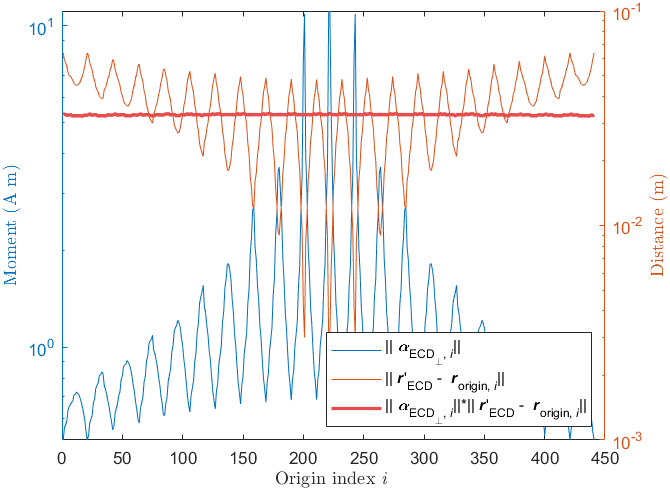}
    \caption{Shows that for all origins that lie in the current triangle source plane, the distance between the origin and ECD fit, $||\mathbf{r}'_{ECD,i} - \mathbf{r}_{origin,i}||$, is inversely proportional to the norm of the ECD moment projected onto the line perpendicular to $\mathbf{r}'_{ECD,i} - \mathbf{r}_{origin,i}$. Since all of these fits have high GOF, it implies there exists a current triangle with two radial edges for each origin that reproduces the reference signal near-equivalently, and each of these near-equivalent triangles have the same area as the original current triangle. }
    \label{fig:6}
\end{figure}

Since only non-radial components of the current have nonzero contribution to the signal, we projected each $\bm{\alpha}_{ECD}'$ onto the line perpendicular to the line between the origin and $\mathbf{r}'_{ECD}$. Let us denote the projected $\bm{\alpha}_{ECD}'$ for the $i$\textsuperscript{th} $xz$ grid origin to be $\bm{\alpha}'_{ECD_{\perp},i}$, the origin position to be $\mathbf{r}_{origin,i}$ , and the ECD fit position to be $\mathbf{r}'_{ECD,i}$ (these are defined relative to the reference origin). We notice that $||\bm{\alpha}'_{ECD_{\perp},i}||$ is approximately inversely proportional to $||\mathbf{r}'_{ECD,i} - \mathbf{r}_{origin,i}||$, as shown in Figure~\ref{fig:6}. Since the GOF is equivalent to the unprojected $\bm{\alpha}_{ECD}'$ case, this implies that there is another near-perfect ECD fit solution, this time oriented tangentially to the observation line with respect to the origin. All of the equivalent triangles having the same surface area guarantees that in the far-field approximation they all converge to the same field of a magnetic dipole with dipole moment $I\mathbf{a}$, where $\mathbf{a}$ is the surface area of the current loop.

\section{Discussion} \label{sec:discussion}

\subsection{Integration path choice defining the magnetic scalar potential}

We have derived an expression for the quasi-static magnetic field in a current-free region that renders radial current segments  magnetically silent. However, this condition holds only because we have defined our magnetic scalar potential with a radial integration path. A radial path has been the convention used in MEG and is likely the most practical to implement, thus we have chosen to present our paper accordingly. In general, if assuming a straight integration path, current segments that are parallel to this integration path or the source-to-field  direction will be magnetically silent. 

A possible application in defining non-radial integration paths may be to suppress the detecting sources in certain orientations, especially if they are not of biological interest. When used in conjunction with multi-dipole fits (discussed in Section~\ref{discussion_coreg} below), there may also be applications in determining an approximate equivalent current loop as discussed in Section~\ref{sec:vol_to_line}. If one defines a few versions of the magnetic scalar potential for an origin, each having a different straight integration path, then current segments of different orientations switch between being magnetically contributing or non-contributing. They will thus, in theory (and with constraints), be able to be localized in an inverse model for at least one magnetic scalar potential definition. Further investigations into these ideas are left for future work.

\subsection{Ambiguous contribution of current segments to the magnetic field}

The Biot-Savart law expresses the magnetic field as a superposition of the contributions of infinitesimally short current segments. Moreover, it indicates that the segments have unique contributions that add up to the total magnetic field by the principle of superposition. However, we have shown in Sections~\ref{sec:theory} and \ref{sec:forward} that in a simply connected source-less region for the more specific electro-quasi-static field case, an alternative formula can be derived in which the contribution of these individual segments is non-invariant with respect to translations of the coordinate system. These contributions differ based on the computational origin since the tangential and radial directions are defined relative to the origin. We have investigated the electro-quasi-static case in detail in this paper since it specifically pertains to the context of MEG and is used for MEG inverse model estimates. Moreover, it allows for the magnetic field to be expanded in terms of vector spherical harmonics \cite{taulu2005presentation,taulu2020unified}, which allows for future analyses in the spatial frequency domain.

In our formula, only non-radial current segments in a current loop contribute to the magnetic field in the electro-quasi-static magnetic forward model.
This has consequences in applications where individual segments of a closed current loop are given specific physical meanings, as their individual contributions may be ambiguous unless the chosen mathematical model is clearly specified. 
For example,  Section~\ref{ecd_results} demonstrates the case of MEG where a primary line current source and the associated passive volume currents are considered. 
In our formulation, motivated by~\cite{sarvas1987basic}, an origin at the center of the spherical head model results in primary current contributions only, i.e., $\mathbf{B}(\mathbf{r}) = \mathbf{b}_{pri}(\mathbf{r})$ and $\mathbf{b}_{vol}(\mathbf{r}) = \mathbf{0}$. However, it is possible to choose an origin that coincides with either of the other two triangle vertices, leading to volume current contributions only, i.e., $\tilde{\mathbf{B}}(\tilde{\mathbf{r}}) = \tilde{\mathbf{b}}_{vol}$ and $\tilde{\mathbf{b}}_{pri}(\tilde{\mathbf{r}}) = \mathbf{0}$. In this translated origin, an ECD fit thus localizes the dipole to the volume current edge with high GOF and a significant LE as seen in Figure~\ref{fig:4}. This may result in an incorrect interpretation of the ECD fit corresponding to the primary current, when it is in fact localized to the volume current edge.

In section~\ref{sec:inverse_relation}, we showed that for origins lying in the plane of the triangle, there is at least one near-perfect ECD fit (or equivalently, current triangle with two radial edges) in the same plane. This is true even if the origin does not coincide with a triangle vertex. Figure~\ref{fig:6} shows that at least one of these near-perfect fits is a current triangle with the same surface area as the original current triangle itself. These ECD fits/equivalent triangles with high GOF that one may obtain when performing an ECD fit do not localize to the primary current edge as desired. As can be seen from its LE subplot in Figure~\ref{fig:5}, they may not necessarily mean anything physically at all. This is another example where prematurely assigning the meaning of a primary current to an ECD fit with high GOF may be misleading. 

\subsection{Co-registration errors in MEG} \label{discussion_coreg}

In practice, physical misinterpretations are especially relevant in scenarios with a co-registration error where the computational origin does not agree with the center of the conductor sphere (reference origin).

From Figure~\ref{fig:5}, we see that compared to the plane parallel to the current triangle, co-registration errors in directions perpendicular to the current triangle result in a faster decay of GOF from the reference origin despite having comparable LE. In fact, as discussed, any co-registration errors in the plane of the current triangle result in high GOF values. This suggests that for an unconstrained ECD fit, one may (with caution) interpret a significant decrease in GOF as a co-registration error off the plane of the triangle. More care may need to be taken to minimize co-registration errors in the plane of the current triangle, since its GOF provides little information about the origin position.

Moreover, short, near-radial and/or deep primary current segments are more prone to ECD fit errors and misinterpretations, since a smaller co-registration error may more easily result in the primary current segment becoming radial and magnetically silent.

The translational invariance property of closed current loop contributions suggest that in any forward or inverse signal calculations, all segments of the loop should be taken into account. The reason for this is that when parametrizing the forward model to include the geometry (position and moment) of a primary current dipole only, the validity of the model depends on the origin of the forward model. Considering the entire loop mitigates co-registration errors  in forward calculations, but in the inverse model, constraints may be needed as mentioned in the introduction.

\begin{figure}[htpb] 
\centering
    \includegraphics[width=0.48\textwidth]{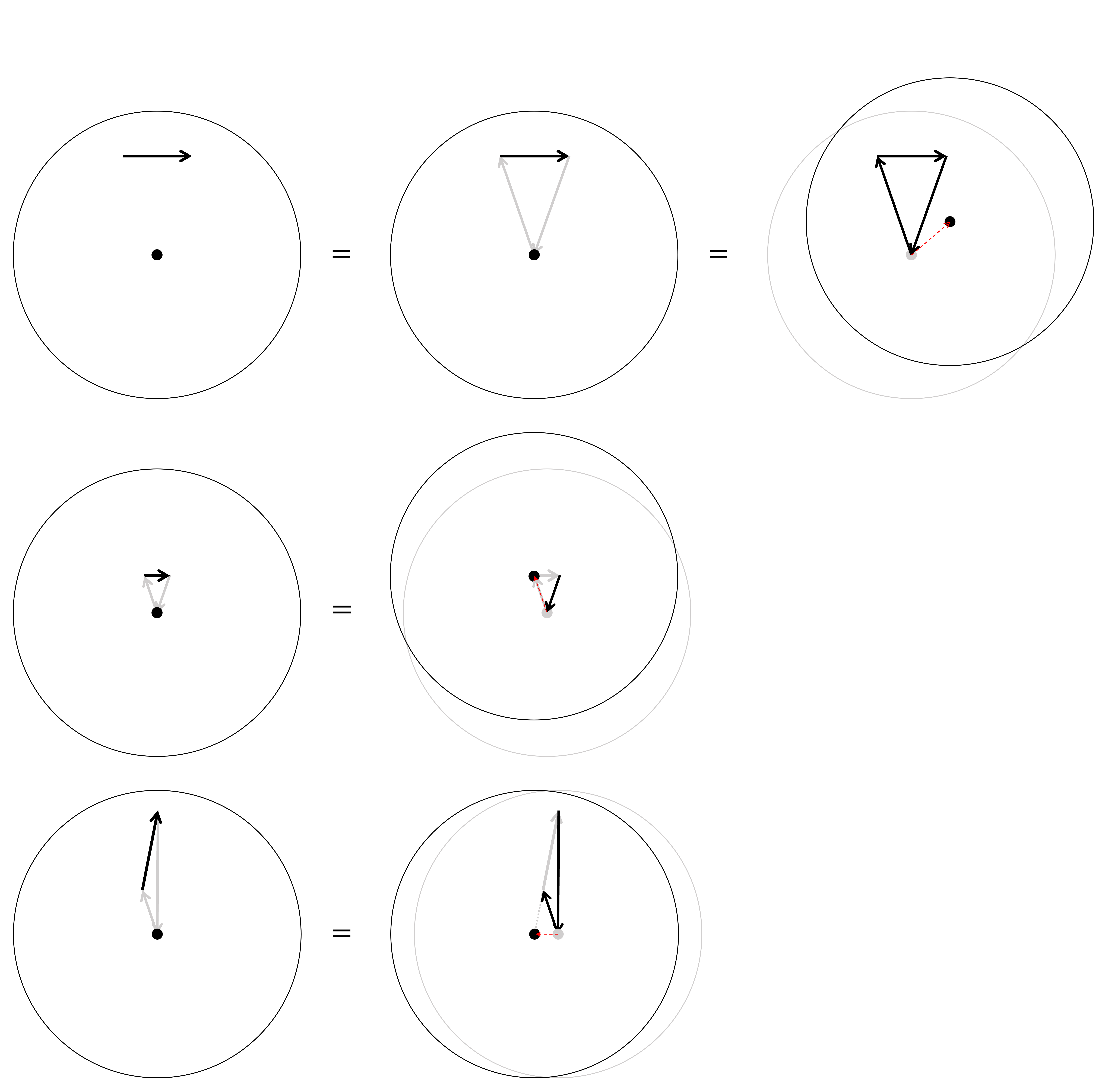}
    \caption{\textbf{First row}: Two magnetically-silent radial volume current edges may be introduced in a spherical head model to form a closed current loop with the primary current. This is equivalent to the case when the origin is translated such that all edges are not radial any longer; all three edges now have non-zero signal contributions. \textbf{Second and third row}: Deep and near-radial current sources are more sensitive to co-registration errors respectively. A small origin translation is sufficient to render them radial and magnetically silent.}
    \label{fig:7}
\end{figure}

In Figure~\ref{fig:7}, we show source configurations that yield equivalent MEG signal contributions in the spherical head case, as well as how short, near-radial and/or deep primary current segments are more sensitive to co-registration errors. In the figure,
black colored arrows represent current segments with non-zero MEG signal contributions, whereas gray colored arrows represent radial/magnetically silent current segments. Black colored dots represent the computational origin in question, and a black colored circle represents the corresponding spherical head model with the black dot located at its center. Gray colored dots and circles represent the original computational origin and corresponding spherical head model used in a previous equality. This figure is an extension of Figure 1.5(d) presented in \cite{hari2017meg} and the figures presented in \cite{ilmoniemi1985forward,ilmoniemi2009triangle}.

However, we note that ECD fits have been used with good success in practice, even in clinical settings (see, e.g., \cite{laohathai2021practical,hari2018ifcn}). Thus, the effects of co-registration errors as described here are likely more relevant with deep sources and tilted dipoles.

\subsection{Equal importance of both primary and volume currents }

It is commonly stated that for a spherical head model, the radial component of the magnetic field has no contribution from volume currents and that radially oriented MEG sensors are only sensitive to the tangential components of the primary current \cite{baillet2001electromagnetic,ilmoniemi2019brain,hamalainen1993magnetoencephalography,ioannides2009magnetoencephalography,ahlfors2019overview,ahlfors2010sensitivity}. The Sarvas approach of obtaining the full magnetic field via only its radial component (shown in Section~\ref{sec:theory}) also leads to a common conclusion that we may be able to reconstruct the entire magnetic field with just the primary current contribution. 

However, in our alternative electro-quasi-static formulation, which is in agreement with the Biot-Savart law, the above statement is true only when the origin is taken to be at the center of the sphere. In fact, at that singular position our formula is equivalent to the Sarvas formula. A shift in origin can render all components of the magnetic field, including the radial component, to have volume current contributions only regardless of the head conducting geometry. As such, we may completely obtain the magnetic field from volume current contributions with specific choices of origins. Thus, the sensitivity of MEG to different components of the underlying current is just a matter of convention used in the forward model.

\subsection{Sensor orientations}

Our results may also be of interest regarding interpreting the significance of sensor orientations. There have been discussions about radially-oriented sensors having higher sensitivity to primary currents and tangentially-oriented sensors having higher sensitivity to volume currents, which is in agreement with the statement above \cite{iivanainen2017measuring,brookes2021theoretical}. However, we have shown that it is possible  to construct formulation in which differently positioned or oriented sensors detecting the magnetic field cannot be considered to be especially sensitive to primary or volume currents. This is again due to the freedom to choose our computational origin to force certain current segments to be radial/magnetically silent or non-radial/contributing.

As for the construction of the sensor array, it has also been shown that a sensor array with diversely-oriented sensors improves the performance of source reconstruction and software shielding capabilities \cite{hochwald1997magnetoencephalography,nurminen2013improving}. It was suggested that the detection of a mixture of both primary and volume contributions with such a sensor array may be a reason behind the improved performance over arrays with purely radially-oriented sensors \cite{hochwald1997magnetoencephalography}. However, following our discussions above, we have shown that it is possible to alter that primary and volume current contributions of a given current configuration independently of sensor orientations, via origin shifts. There is thus room for further work to investigate the priority of sensors and their orientations for inverse modeling with respect to origin choice.

Note that all the above discussions about the effects of origin choice in the context of MEG calculations do not hold for electroencephalography (EEG), since EEG measures the electric scalar potential, which is invariant under coordinate translations.

\section{Conclusion} \label{sec:conclusion}

In this paper, we have shown that in the electro-quasi-static magnetic forward problem within a source-free region, the translational invariance of magnetic scalar potentials and magnetic fields holds only for closed current loops, not open current segments. Compact formulas were derived for the magnetic scalar potential and magnetic field that indicate that by defining the magnetic scalar potential with a radial integration path, radially-oriented current segments with respect to the origin in each coordinate system are magnetically silent. This simplifies source considerations needed for typical magnetic field calculations using the Biot-Savart law. On the other hand, electric scalar potentials and electric fields produced within any general region by any constant current segments (both open or closed) in the magneto-quasi-static electric forward problem are always translationally invariant. 

In the context of MEG where focal sources may be modeled as triangle current loops, the non-uniqueness of open current segment contributions to the signal in translated coordinate systems results in the LE and GOF for ECD fits exhibiting different behaviours relative to the co-registration error. The LE increases as the origin choice distance increases from the reference origin as expected, however the GOF is high when the origin choice lies on the same plane as the current triangle, especially around the region of the triangle vertices. GOF decays quickly for co-registration errors away from the current triangle plane. Hence, in general, GOF does not accurately indicate LE or co-registration error behaviors or amounts.

The sensitivity of MEG to primary vs. volume currents is thus a function of the computational model. Although the model can be parametrized to only refer to the true primary current source (equivalent dipole) when the origin is at the center of a spherical conductor, in the same geometry a similar model can be found that only has parameters related to the volume current when the origin is shifted to a particular off-center location. 
The translational non-invariance of the signal contributions from each segment of a closed current loop suggests that a simple shift in origin can determine the amount of primary and volume current contributions. Since only the closed loop contribution is guaranteed to be translationally invariant, it is preferable that the entire loop is taken into account (i.e. both primary and volume currents) for both forward and inverse methods and  Biot-Savart-type calculation is applied.

\appendix

\section{Expressions in the far-field approximation} \label{sec:appendix_dipolecheck}

Here, we show that \eqref{eq:novel_U} and \eqref{eq:novel_B} reduce to the expected magnetic dipole form in the far-field region.

First, note that the vector area $\mathbf{K}$ is 
\begin{align}
    \mathbf{K} = -\frac{1}{2I} \int_\mathcal{C} d\mathbf{L}'
\end{align}
and that the magnetic dipole moment is $\mathbf{m} = I\mathbf{K}$. When $\abs{\mathbf{r}} >> \abs{\mathbf{r}'}$, we see that \eqref{F} and \eqref{grad_F} are approximately
\begin{align}
    F(\mathbf{r},\mathbf{r}') &\approx 2r^3 \\
    \grad F(\mathbf{r},\mathbf{r}') &\approx 6r\mathbf{r}.
\end{align}
By substituting the above in, it is easy to see that the magnetic scalar potential \eqref{eq:novel_U} and magnetic field \eqref{eq:novel_B} in the far-field region are approximately
\begin{align}
    U(\mathbf{r}) \approx \frac{1}{4 \pi} \frac{\mathbf{m} \cdot \hat{\mathbf{r}}}{r^2} 
\end{align}
and
\begin{align}
    \mathbf{B} (\mathbf{r}) \approx - \frac{\mu_0}{4\pi} \frac{1}{r^3}\left[ 3(\mathbf{m} \cdot \hat{\mathbf{r}}) \hat{\mathbf{r}} - \mathbf{m} \right]
\end{align}
respectively, as desired.

\section{Equivalent representations of source currents}

\subsection{Equivalent line current for volume currents} \label{sec:volume_current_line}

By Kirchoff's junction rule, the volume current may be equivalently represented as a finite sum of $P$ line current contributions, each flowing from the source to the sink. The $p$\textsuperscript{th} line current component with current $I_p$ can be parametrized from $t \in [0,1]$ as $\mathbf{r}'_{p}(t)$, with $\mathbf{h}'_p(t)$ as the vector that points in the tangential direction of the line segment. From the volume current equivalent of \eqref{U_pri} (which is the precursor to \eqref{U_vol} before Stokes' theorem was performed), the contribution of the volume current towards the magnetic scalar potential may thus be written as
\begin{equation} 
    u_{vol}(\mathbf{r}) = -\frac{1}{4\pi} \sum_{p=1}^P I_p q_p(\mathbf{r}) \label{eq:uvol_line_sum}
\end{equation}
where 
\begin{equation}
    q_p(\mathbf{r}) = \int_0^1 \frac{\mathbf{r}'_p(t) \times \mathbf{r} \cdot \mathbf{h}'_p(t)}{F(\mathbf{r},\mathbf{r}_p'(t))} dt
\end{equation}
and $\sum_{p=1}^P I_p = I$ necessarily.

We show that the sum over all $P$ line integrals in equation \eqref{eq:uvol_line_sum} can be reduced to one equivalent line integral with current $I$ for some field point $\mathbf{r} = \mathbf{r}_a$ as follows.

First, let us consider two distinct line segments $\mathbf{r}_{1}'(t)$ and $\mathbf{r}_{2}'(t)$ that connect from the source to the sink, with currents $I_1$ and $I_2$ respectively, as well as their corresponding line integrals $q_1(\mathbf{r}_a)$ and $q_2(\mathbf{r}_a)$. We may consider a parametrization of lines from $w \in [0,1]$ that sweep out an open surface $\mathbf{s}'(t,w)$ with $\mathbf{r}_{1}'$ and $\mathbf{r}_{2}'$ forming its boundary; i.e., $\mathbf{s}'(t,0) = \mathbf{r}_{1}'(t)$ and $\mathbf{s}'(t,1) = \mathbf{r}_{2}'(t)$. The intermediate value theorem tells us that there must exist an equivalent $w = w_{eq,a}$, i.e. a path $\mathbf{s}'(t,w_{eq,a}) = \mathbf{r}'_{eq,a}(t)$ that has tangential direction $\mathbf{h}'_{eq,a}(t)$, such that
\begin{align}
    q_{eq,a}(\mathbf{r}_a) &\equiv \int_0^1 \frac{\mathbf{r}'_{eq,a}(t) \times \mathbf{r}_a \cdot \mathbf{h}'_{eq,a}(t)}{F(\mathbf{r}_a,\mathbf{r}'_{eq,a}(t))} dt \\
    &= \frac{I_1 q_1(\mathbf{r}_a)+I_2 q_2(\mathbf{r}_a)}{I_1 + I_2} \\
    \implies (I_1 + &I_2) q_{eq,a}(\mathbf{r}_a) = I_1 q_1(\mathbf{r}_a) + I_2 q_2(\mathbf{r}_a). \label{eq:ivt_res}
\end{align}
Note that the subscript notation ``\textit{eq,a}'' denotes a quantity of the equivalent path when evaluated at $\mathbf{r}_a$. Equation \eqref{eq:ivt_res} tells us that when evaluated at the field point $\mathbf{r}_a$, the magnetic contribution from two line currents with the same source and sink can always be equivalently expressed with a single line current. With this result, we may collapse the sum in \eqref{eq:uvol_line_sum} in a pair-wise manner to obtain an equivalent line current with current density $I$. Volume currents may therefore always be represented by a simple line current that forms a closed loop with the primary current for each field point.

However, it appears that it may be nontrivial to show if a common equivalent current loop holds for all field points; i.e. if there exists a line current such that the statement
\begin{align}
    (I_1 + I_2) q_{eq,a} (\mathbf{r}_b) = I_1 q_1 (\mathbf{r}_b) + I_2 q_2 (\mathbf{r}_b)
\end{align}
holds true for $\mathbf{r}_b \neq \mathbf{r}_a$. This investigation is a potential topic for a future study.


\subsection{Equivalent electric current dipole for a straight line current} \label{sec:mean_value_theorem}

Let us consider a current source that is an open straight line segment with length $l$ and current strength $I$; i.e., $\mathbf{J} dv' = I d\mathbf{l}'$. This segment can be parametrized as $\mathbf{r}' (t)= \mathbf{r}'_0 + \mathbf{h}' t$, $t \in [0,l]$, where $\mathbf{r}'_0$ is the position of the starting end of the segment and $\mathbf{h}'$ is the unit vector pointing in the direction of current flow. Let us also again consider some field point $\mathbf{r} = \mathbf{r}_a$. By \eqref{biot}, the magnetic field produced by this current source evaluated at $\mathbf{r}_a$ may be written in a form mathematically similar to \eqref{U_pre}, 
\begin{align}
    \mathbf{b} ( \mathbf{r}_a) = \frac{\mu_0 I}{4 \pi} \left[\mathbf{h}' \times \left(\mathbf{r}_a - \mathbf{r}_0' \right)\right] \int_0^l \frac{dt}{\abs{\mathbf{r}_a - \mathbf{r}_0' - \mathbf{h}' t}^3}. \label{B_parametrized}
\end{align}
By the mean value theorem for integrals with scalar field integrands, we know that there must exist a $t = t_D \in [0,l]$ such that \eqref{B_parametrized} is equivalently
\begin{align}
    \mathbf{b} ( \mathbf{r}_a) =  \frac{\mu_0 Il}{4 \pi}  \frac{\mathbf{h}' \times \left(\mathbf{r}_a - \mathbf{r}_0' \right)}{\abs{\mathbf{r}_a - \mathbf{r}_0' - \mathbf{h}' t_D}^3}.
\end{align}
This is the form of the magnetic field produced by a dipole, given by \eqref{Bpri_dip} with $N=1$, $\mathbf{D} = Il \mathbf{h}'$, and $\mathbf{r}_D = \mathbf{r}_0' + \mathbf{h}' t_D$. This means that there is a guaranteed ECD solution of a dipole with strength $Il$ pointing in the direction of the straight line current segment that produces an identical signal as the line segment itself when evaluated at $\mathbf{r}_a$.

However, as in the section before, it appears that it may be nontrivial to show if there is a common equivalent dipole for the straight line current that holds for all field points, i.e. if 
\begin{align}
    \int_0^l \frac{dt}{\abs{\mathbf{r}_b - \mathbf{r}_0' - \mathbf{h}' t}^3} \ dt = \frac{1}{\abs{\mathbf{r}_b - \mathbf{r}_0' - \mathbf{h}' t_D}^3}.
\end{align}
holds true for $\mathbf{r}_b \neq \mathbf{r}_a$. Again, this investigation is a potential topic for a future study.

\section{The magneto-quasi-static electric forward problem} \label{sec:appendix_Efield}

Here, we reaffirm that the electric scalar potential and electric fields produced by open line charges with constant charge density are translationally invariant.

Coulomb's law states that the electric field of a line charge with charge density $\lambda ( \mathbf{r}')$ along a line segment $\mathcal{D}$ is given by
\begin{align}
      \mathbf{E} (\mathbf{r}) = \frac{1}{4 \pi \epsilon_0} \int_\mathcal{D} \frac{\lambda(\mathbf{r}') (\mathbf{r} - \mathbf{r}')}{\abs{\mathbf{r} - \mathbf{r}'}^3} \ dl'
\end{align}
where $\epsilon_0$ is the vacuum permittivity. Let us assume a straight line charge of length $l$ with constant charge density $\lambda(\mathbf{r}') = \lambda$. Similar to Section~\ref{sec:mean_value_theorem}, this charge segment can be parametrized as $\mathbf{r}' = \mathbf{r}'_0 + \mathbf{h}' t'$, $t' \in [0,l]$, where $\mathbf{r}'_0$ is the position of one end of the line segment and $\mathbf{h}'$ is the vector pointing towards the other end. Its electric field contribution can thus also be written as
\begin{align}
    \mathbf{E} ( \mathbf{r}) &= \frac{\lambda}{4 \pi \epsilon_0} \int_0^l \frac{\mathbf{r} - \mathbf{r}'_0 - \mathbf{h}' t'}{\abs{\mathbf{r} - \mathbf{r}'_0 - \mathbf{h}' t'}^3} \ dt'.
\end{align}
Under magneto-quasi-static approximation, we have $\grad \times \mathbf{E} (\mathbf{r}) = \mathbf{0}$ and $\mathbf{E} (\mathbf{r}) = - \grad V (\mathbf{r})$ as stated before, thus we may find the electric scalar potential via a similar approach as presented in Section~\ref{sec:theory},
\begin{align}
    V (\mathbf{r}) &= - \int_0^\infty \grad V (\mathbf{r} + t \mathbf{e}_r) \cdot \mathbf{e}_r \ dt \\
    &= \int_0^\infty \mathbf{E} (\mathbf{r} + t \mathbf{e}_r) \cdot \mathbf{e}_r \ dt
\end{align}

Let $\mathbf{a}_E = \mathbf{r} - \mathbf{r}'_0 - \mathbf{h}' t'$, $k_E = \mathbf{a}_E \cdot \mathbf{e}_r$ and $y_E = t + k_E$. Then, we have
\begin{align}
    V (\mathbf{r}) &= \frac{\lambda}{4 \pi \epsilon_0}  \left.\Bigg( \int_0^l k \int_0^\infty \frac{1}{\abs{\mathbf{a}_E + \mathbf{e}_r}^3} \ dt dt'  \right. \nonumber \\
    &\hspace{5mm}+ \left. \int_0^l \int_0^\infty \frac{t}{\abs{\mathbf{a}_E + t \mathbf{e}_r}^3} \ dt dt' \right.\Bigg) \label{V_pre}
\end{align}
The inner integral in the first line of \eqref{V_pre} is given by \eqref{integral_result}, whereas the inner integral of the second line is given by a similar result, 
\begin{align}
    \int_0^\infty \frac{t}{\abs{\mathbf{a}_E + t \mathbf{e}_r}^3} \ dt = \frac{1}{a_E+k}.
\end{align}
Thus, we have
\begin{align}
    V (\mathbf{r}) = \frac{\lambda}{4 \pi \epsilon_0} \int_0^l \frac{1}{a_E} \ dt'.
\end{align}
Since $a_E$ is independent of origin choice, this expression indicates that the electric scalar potential of a straight line charge (and hence the electric field as well) is translationally invariant. Note that there is no condition that the region of interest must be source-less. This result is expected to be consistent with the result in the current loop magnetic field counterpart; individual point charges are analogous to current loops (magnetic dipoles) since they are the the fundamental electric and magnetic sources respectively. 

For completeness, we state the expression when the integral has been evaluated. Let  $\mathbf{b} = \mathbf{r} - \mathbf{r}'_0$, $c = \mathbf{b} \cdot \mathbf{h}'$, and $z = t' - c$ so $dz = dt'$. Then, we have
\begin{align}
    V(\mathbf{r}) 
    = \frac{\lambda}{8 \pi \epsilon_0} \ln \left( \frac{\sqrt{l^2 - 2lc + b^2} + (l-c)}{\sqrt{l^2 - 2lc + b^2} - (l-c)} \cdot \frac{b+c}{b-c} \right).
\end{align}
This expression is presented in many electromagnetism textbooks, although likely with different parameters. For instance, see Problem 2.25 in \cite{griffiths}.

\section{An equivalent form for the \texorpdfstring{$U(\mathbf{r})$}{Ur} and \texorpdfstring{$\mathbf{B}(\mathbf{r})$}{Br} line integrals} \label{sec:appendix_greens_thm}

Here, we show that for a flat loop with origin in the same plane as the loop, there is an equivalent line integral form of \eqref{eq:novel_U} and \eqref{eq:novel_B} with respect to outward pointing normal of the loop, instead of along the tangential direction of the loop. 

Let us denote the area unit normal vector of the flat loop to be $\mathbf{n}_A$. Green's theorem tells us that \eqref{eq:novel_U} may be written as 
\begin{align}
     U(\mathbf{r}) = \frac{I}{4\pi} \iint_{A_\mathcal{C}} \grad' \times \left[ \frac{\mathbf{r} \times \mathbf{r}'}{F(\mathbf{r},\mathbf{r}')} \right] \cdot \mathbf{n}_A \ dA
\end{align}
where $\grad'$ acts only on the source primed coordinates. Using the vector identity $\grad'(\mathbf{x} \times \mathbf{y}) = \mathbf{x}(\grad' \cdot \mathbf{y}) -  \mathbf{y}(\grad' \cdot \mathbf{x}) + (\mathbf{y} \cdot \grad')\mathbf{x} - (\mathbf{x} \cdot \grad')\mathbf{y}$ with $\mathbf{x} = \mathbf{r}$ and $\mathbf{y} = \mathbf{r}'/F(\mathbf{r},\mathbf{r}')$, we have 
\begin{align}
     U(\mathbf{r})_{A_\mathcal{C}} = \frac{I}{4\pi} \iint \left\{ (\mathbf{r} \cdot \mathbf{n}_A) \grad' \cdot \left(\frac{\mathbf{r}'}{F(\mathbf{r},\mathbf{r}')} \right) \right. \nonumber \\
     \left. - \left[(\mathbf{r} \cdot \grad') \frac{\mathbf{r}'}{F(\mathbf{r},\mathbf{r}')}\right] \cdot \mathbf{n}_A     \right\} \ dA. 
\end{align}
If the origin is in the plane of the loop, i.e. if $\mathbf{n}_A \cdot \mathbf{r}' = 0$, then the second term in the integrand vanishes, and the above becomes
\begin{align}
     U(\mathbf{r}) = \frac{I}{4\pi} (\mathbf{r} \cdot \mathbf{n}_A) \iint \grad' \cdot \left(\frac{\mathbf{r}'}{F(\mathbf{r},\mathbf{r}')} \right) \ dA. \label{eq:U_area}
\end{align}
The 2D divergence theorem tells us that this may again be reduced into a line integral 
\begin{align}
     U(\mathbf{r}) = \frac{I}{4\pi} (\mathbf{r} \cdot \mathbf{n}_A) \int_\mathcal{C} \frac{\mathbf{r}'}{F(\mathbf{r},\mathbf{r}')} \cdot \mathbf{n}_l \ dl'
\end{align}
where $\mathbf{n}_l$ is the outward pointing normal along the loop. The magnetic field is then
\begin{align}
    \mathbf{B}(\mathbf{r}) = - \frac{\mu_0 I}{4\pi} \mathbf{n}_A \cdot \int_\mathcal{C} (\mathbf{r}' \cdot \mathbf{n}_l) \grad \left(\frac{\mathbf{r}}{F(\mathbf{r},\mathbf{r}')}\right) dl'
\end{align}
where the gradient in the integrand is given by \eqref{eq:grad_r/F}.

Note that \eqref{eq:U_area} implies that a loop-preserving transformation of the parametrization $\mathbf{r}' \rightarrow \mathbf{r}' + F(\mathbf{r},\mathbf{r}') \grad \times \mathbf{z}$ for any arbitrary vector field $\mathbf{z}$ yields an invariant magnetic scalar potential $U(\mathbf{r})$. This is in agreement with the fact that there are more than one current distributions that produce the same magnetic field, as discussed in the introduction.

\bibliography{references.bib}

\end{document}